\documentclass[10pt]{iopart}

\usepackage[utf8]{inputenc}
\usepackage[T1]{fontenc}
\usepackage[english]{babel}
\usepackage[colorlinks,citecolor=red,urlcolor=blue,bookmarks=false,hypertexnames=true]{hyperref}
\usepackage{pdfpages}
\usepackage{amsthm}
\usepackage{cite}
\usepackage{iopams}

\expandafter\let\csname equation*\endcsname\relax
\expandafter\let\csname endequation*\endcsname\relax
\expandafter\let\csname subequations*\endcsname\relax
\expandafter\let\csname endsubequations*\endcsname\relax

\usepackage{amsmath, amssymb, esint}
\usepackage{cases}
\usepackage{accents}
\usepackage{mathtools}
\usepackage{tensor}
\usepackage[math]{blindtext}
\usepackage[margin=2.5cm]{geometry}
\usepackage{stackrel}
\usepackage{citesort}
 
\def\divider{\par
  \vskip 0.5em
  \centerline{\hbox to 0.5\hsize{\hrulefill}}
  \vskip 0.5em
}

\begin{document}

\title{Superspace worldline formalism approach to Quantum Gravity: dimensional reduction and Holography}

\author{J-B. Roux$^\ast$}
\address{$^\ast$ University of Aix-Marseille, \textsc{France} (Now unaffiliated, independent researcher)}
\ead{jeanbaptiste.roux@live.fr}
\begin{minipage}{\textwidth}
\begin{abstract}
    Using the ADM formalism, we demonstrate that the Hamiltonian formulation of Quantum Gravity is exactly in the form of a worldline (WL) formalism in the superspace. We then show that the Keldysh partition function reduces to the partition function of Euclidean 3D gravity. After discussing the meaning of the time parameter, we show that in the gauge fields formalism, our Keldysh partition function reduces to a generating functional of a 2D Conformal Field Theory (CFT). This functional exhibits a minimal time-lapse proportional to the square root of the cosmological constant. From the viewpoint of the Chern-Simons/Liouville correspondence, we calculate the exchange of virtual gravitons between two massive probes from the cosmological boundary.\\
    \noindent{\textsc{keywords}--- \it{Worldline formalism, Quantum Gravity, AdS/CFT}\,\text{|}}
    \submitto{\CQG}
\end{abstract}
\end{minipage}

\section*{Introduction}

The quest for Quantum Gravity has led physicists to develop various and seemingly drastically different approaches to unify General Relativity and Quantum Mechanics. Examples include Super String Theory, Loop Quantum Gravity, Asymptotic Safety, Non-commutative Geometry, Causal Dynamical Triangulation, and others \cite{oriti_2009}. In fact, they can share common properties, like (in some sense) the ``discreteness'' of space-time \cite{Hossen_2013} and an effective dimension of $\sim 2$, rather than the four usual dimensions \cite{Carlip_2017}. Our approach exhibits a minimum time parameter as well as a minimum space interval parameter and a dimensional reduction from 3+1 dimensions to two dimensions. This approach recognizes the Hamiltonian path integral of General Relativity as nothing more than a worldline formalism in the superspace of spatial metrics. Moreover, it seems that when switching to the gauge connections formalism, one encounters a 2D CFT, namely a Wess-Zumino-Witten (WZW) model, as the underlying theory, which can be further reduced to a Liouville gravity theory.\\ 
\indent First, we briefly recall the basis of the ADM formalism \cite{Arnowitt_2008,corichi2022introduction,gourgoulhon200731}, its first order counterpart \cite{Alexandrov_1998,Alexandrov_2000} and worldline (WL) formalism \cite{edwards2019quantum,corradini2021spinning}. The goal of the first part is to introduce the basics of these concepts rather than conducting a thorough review of them. Second, we use the Hamiltonian path integral of the theory (the partition function). We solve the diffeomorphism constraint by confining it into a closure relation of generalized Kodama states and recognizing that the Keldysh partition function (as well as the regular thermal partition function) automatically satisfies it when the lapse function is constant. Then, we give the expression of the Keldysh partition function with a fixed boundary expressed as a 3D Euclidean partition function. To make the path integral invariant by gauge parameter choice, we use a particular choice of hypersurfaces over time, which are no longer space-like. Next, upon combining the triads and the spatial spin connection into gauge fields to describe hypersurfaces, we find that the Keldysh partition function is reduced to a 2D WZW model. Identifying the level of the theory with an integer value to obtain a large gauge transformation invariance, we find a discrete time. Finally, we analyze the 2D theory from the viewpoint of the AdS/CFT correspondence and conclude that if we include massive probes in the theory, then there are gravitational imprints of their presence on the cosmological boundary, that we calculate, in the form of virtual graviton exchange. A key property of the WZW model \cite{Nair3DSchro,EberhardtWZW} is given in the appendix.

\section*{I. Preliminary reminders}

\subsection*{1) ADM formalism}

The Arnowitt-Deser-Misner formalism consists of a foliation of space-time by space-like hypersurfaces \cite{Arnowitt_2008} as a 3+1 decomposition. Formally, we can say that a space-time $\mathcal{M}$ is the continuous union of the space-like hypersurfaces $\Sigma_t$, where $t$ is a parameter representing their label in time:
\begin{equation}\label{eqa1}
    \mathcal{M} = \bigcup_{t\in I}\Sigma_t
\end{equation}
The key to the ADM formalism is the decomposition of the metric into three parts: the lapse function $N$, the shift vector $N_i$, and the spatial metric $\gamma_{ij}$. In the following, we will always use $i,j,k,l$ for the spatial indices of the tensors and $\mu, \nu$ for the space-time indices (although these are quite absent from the core of this work). Specifically, the lapse function is simply $\sqrt{-1/g^{00}}$, and the shift vector is to be interpreted as ``how a point is displaced in space through time along a geodesic''. We can express the four-dimensional metric and its inverse with these quantities as follows:
\begin{equation}\label{eqa2}
    g_{\mu \nu} = \left( \begin{matrix} -N^2+N_i N^i & N_j \\ N_i & \gamma_{ij} \end{matrix} \right),\,\,\, g^{\mu \nu} = \left( \begin{matrix} -\frac{1}{N^2} & \frac{N^j}{N^2} \\ \frac{N^i}{N^2} & \gamma^{ij}-\frac{N^i N^j}{N^2} \end{matrix} \right)
\end{equation}
Therefore, the determinant of the metric $g$ is $\sqrt{-g} = N \sqrt{\gamma}$. Now, suppose that $I$ in (\ref{eqa1}) is an interval, then we can construct a 1-form $dn = -Ndt$ pointing outward from the foliation. We define the extrinsic curvature (the second fundamental form) as $K_{\mu \nu}=-\perp (\nabla_\mu n_\nu) \equiv -\gamma^\rho_\mu \gamma^\sigma_\nu \nabla_\rho n_\sigma$. Similarly, we define its trace as, $K=\epsilon \nabla_\mu n^\mu$. The definition of the extrinsic curvature ensures that it lives on the hypersurface ($\epsilon = +1$ for space-like $n$ and $-1$ for time-like $n$). Indeed, only the spatial components of $\gamma_{\mu \nu}$ are non-zero, and this quantity, also known as the induced metric or first fundamental form is a projector onto the hypersurface. We can gather some identities known as the Gauss-Codazzi and Ricci equations which we will not prove (see \cite{gourgoulhon200731}):
\begin{align}
    \perp R_{\lambda \mu \sigma \nu} =& {}^{(3)}R_{\lambda \mu \sigma \nu}+K_{\lambda \sigma} K_{\mu \nu}-K_{\mu \sigma}K_{\nu \lambda}\label{eqa3}
\\
    \perp R_{\textbf{n} \mu \sigma \nu} =& -\mathcal{D}_\nu K_{\mu \sigma}+\mathcal{D}_\sigma K_{\mu \nu}\label{eqa4}
\\
    \perp R_{\textbf{n} \mu \textbf{n} \nu} =& \pounds_\textbf{n} K_{\mu \nu}+K^\lambda_\mu K_{\lambda \nu} + \frac{1}{N}\mathcal{D}_\mu \mathcal{D}_\nu N\label{eqa5}
\\
    \perp R_{\mu \nu} =& {}^{(3)}R_{\mu \nu} - \frac{1}{N}\mathcal{D}_\mu \mathcal{D}_\nu N - \pounds_\textbf{n} K_{\mu \nu}+KK_{\mu \nu}-2K^\lambda_\mu K_{\lambda \nu}\label{eqa6}
\\
    \perp R_{\textbf{n} \nu} =& -\mathcal{D}^\mu K_{\mu \nu}+\mathcal{D}_\nu K\label{eqa7}
\\
    R_{\textbf{n} \textbf{n}} =& \pounds_\textbf{n} K-K^{\mu \nu} K_{\mu \nu} + \frac{1}{N}\mathcal{D}_\mu \mathcal{D}^\mu N\label{eqa8}
\\
    R =& {}^{(3)}R+K^2+K^{\mu \nu}K_{\mu \nu}-2 \pounds_\textbf{n} K-\frac{2}{N}\mathcal{D}_\mu \mathcal{D}^\mu N\label{eqa9}
\end{align}
Where $T_\textbf{n} \equiv n^\mu T_\mu$, $\pounds_\textbf{n}$ is the Lie derivative along $n$ and the ``acceleration'' is defined as $a^\mu \equiv \nabla_\textbf{n} n^\mu = \mathcal{D}^\mu \ln N$ (with $\mathcal{D}$ the spatial covariant derivative). Equations (\ref{eqa3}), (\ref{eqa6}), and (\ref{eqa9}) are the Gauss identities (Gauss, contracted Gauss, and scalar Gauss respectively), whereas (\ref{eqa4}) and (\ref{eqa7}) are the Codazzi and contracted Codazzi respectively. More precisely, the scalar Gauss relation is a combination of Equations (\ref{eqa8}) and (\ref{eqa9}). The latter is interesting, as it is well known that the Einstein-Hilbert action is an integral over the entire space-time of the Ricci scalar. We obtain the Einstein-Hilbert action without a cosmological constant as:
\begin{align}\label{eqa10}
    S_\text{E.H.}[g] =& \frac{1}{2\kappa}\int_{\mathcal{M}}d^4x\sqrt{-g}R
\nonumber \\
    =& \frac{1}{2\kappa}\int_{\mathcal{M}}d^4x N\sqrt{\gamma}\left( {}^{(3)}R+K^2+K^{\mu \nu}K_{\mu \nu} -2 \partial_\textbf{n} K\right)-\frac{1}{2\kappa}\int_{\mathcal{M}} d^4x \sqrt{\gamma} 2 \mathcal{D}_\mu \mathcal{D}^\mu N
\nonumber \\
    \stackrel{!}{=}& \frac{1}{2\kappa}\int_{\mathcal{M}}d^4x N\sqrt{\gamma}\left( {}^{(3)}R-K^2+K^{\mu \nu}K_{\mu \nu}\right)+\frac{1}{\kappa}\int_{\partial \mathcal{M}}d^3x \sqrt{\gamma} (\epsilon K+ n_\mu \mathcal{D}^\mu N)
\end{align}
These notations are not very enlightening because, in our use of the Stokes theorem for the last equality, this is not a simple integration by part. It is an equality between the integral over $\mathcal{M}$ of a closed 4-form $d\omega$, and the integral over $\partial \mathcal{M}$ of the 3-form $\omega$. This explains the disappearance of the $N$ factor in the boundary integral. Thus, adding the Gibbons-Hawking-York (GHY) term partially eliminates the boundary integral. Indeed, if for $J\subseteq I$ there is a boundary on each of the space-like hypersurfaces of the family $(\Sigma_t)_{t \in J}$, then $\partial \mathcal{M}$ has space-like boundaries in addition to the two time-like boundaries. The Hamiltonian density of the theory is then the Legendre transform $H = \pi^{ij}\partial_t \gamma_{ij}-L$, where $\pi^{ij} = \frac{\partial L}{\partial \partial_t \gamma_{ij}} = \sqrt{\gamma}(K\gamma^{ij}-K^{ij})$. It can be found \cite{gourgoulhon200731} that the bulk Hamiltonian is of the form:
\begin{equation}\label{eqa11}
    H=N\mathcal{H}+N_i \mathcal{P}^i+2\partial_j \left( \pi^{ji}N_i-\frac{1}{2}\pi N^j \right)
\end{equation}
Where $\mathcal{H}=G_{ijkl}\pi^{ij}\pi^{kl}-\sqrt{\gamma}{}^{(3)}R$ is the Hamiltonian constraint and $\mathcal{P}^i=-2\mathcal{D}_j \pi^{ij}$ is the momentum constraint. Here, $G$ is the DeWitt supermetric $G_{ijkl} = \frac{1}{2\sqrt{\gamma}}(\gamma_{ik}\gamma_{jl}+\gamma_{il}\gamma_{jk}-\gamma_{ij}\gamma_{kl})$. By taking the reverse Legendre transform one can find the Hamiltonian action of the theory, with the GHY term and its counter-term:
\begin{align}\label{eqa12}
    S[g,\pi] =& \frac{1}{2\kappa}\int_0^{\beta_m} dt \int_\Sigma d^3x\left( \pi^{ij}\partial_t \gamma_{ij}-N\mathcal{H}-N_i \mathcal{P}^i \right)-\frac{1}{\kappa}\int_{\partial \mathcal{M}} d^3y\,n_i\left( N_j \left(\pi^{ij} -\frac{1}{2}\gamma^{ij}\pi \right) + \sqrt{\gamma}\mathcal{D}^i N\right)
\nonumber \\    
    &+\frac{1}{\kappa}\int_0^{\beta_m}dt\int_{\partial \Sigma} d^2y N\sqrt{\sigma}(K-K_0)
\end{align}
where $\beta_m$ is the measure of $I$ in equation (\ref{eqa1}). Because $n$ is perpendicular to the boundary, we conclude that the first boundary term vanishes identically on the time-like boundaries. For space-like boundary terms, we impose boundary conditions $\pi|_{\partial \Sigma}=0$ and $n^i|_{\partial \Sigma}=0$. Thus, the Hamiltonian action is simply:
\begin{align}\label{eqa13}
    S[g,\pi] =& \frac{1}{2\kappa}\int_0^{\beta_m} dt \int_\Sigma d^3x\left( \pi^{ij}\partial_t \gamma_{ij}-N\mathcal{H}-N_i \mathcal{P}^i \right)+\frac{1}{\kappa}\int_0^{\beta_m}dt\int_{\partial \Sigma} d^2y N\sqrt{\sigma}(K-K_0)
\end{align}
Note that lapse function $N$ and shift vector $N_i$ are both Lagrange multipliers. This action is used in the following. We choose the Hamiltonian action formalism because the Feynman path integral comes from the phase-space path integral in the first place. Thus, the latter can be considered more fundamental than the former.\\
\indent In (\ref{eqa13}), $\Sigma$ is in general dependent on the time parameter $t$, as shown in (\ref{eqa1}). However, if we look at $\Sigma$ as being a topological space, as for the integration over a chain, we can drop this label because throughout, we assume no change in topology over time.

\subsection*{2) Changing to first-order formalism}

Now that we have derived the Hamiltonian action (\ref{eqa13}) of General Relativity in the second-order formalism, we want to have its first-order counterpart, the Palatini action. To simplify a bit the calculations, and because in part II we will work with fixed boundaries, we omit in the first place the boundary terms. In part II. 1) we will focus on a peculiar formulation of the first-order gravity, namely the Holst action. Our work is not based on this action \textit{per se}, only on its non-topological part which coincides with the Palatini action. But the study of the Holst action has already been carried out in \cite{Alexandrov_1998, Alexandrov_2000}, and it is easy to recover the Palatini theory from the Holst one. The main difference between second- and first-order formalisms is the appearance of second-class constraints, for the latter. Let us introduce in the first place a complex gauge field $A^a_i \in \mathbb{C}$:
\begin{align}\label{eqa14-15-16.0.1}
    \sqrt{\frac{3}{\Lambda}} A^0_4 =& Ndt+\chi_a E^a_i dx^i,\,\,\,\sqrt{\frac{3}{\Lambda}}A^a_4 = E^a_i dx^i+N^i E^a_i dt
\\
    A^a_i =& \epsilon^{abc} A_{bc \, i},\,\,\, A^a_0 = \epsilon^{abc} A_{bc\,0}
\\
    \mathcal{F}^a_{ij} =& \epsilon^{abc}\mathcal{F}_{bc\,ij}
\end{align}
Where $\eta_{ab}E^a_i E^b_j = \gamma_{ij}$, and $\mathcal{F}{}^a_{ij}$ is the curvature of $A^a_i$. Furthermore, one introduces the notations $\undertilde{N} = \sqrt{|\gamma|}^{-1} N$ and $\widetilde{E}{}^a_i = \sqrt{|\gamma|}E^a_i$ to simplify the calculations. Then, the Holst action with imaginary Barbero-Immirzi parameter (which we will call ``the Ashtekar'' action, following \cite{Alexandrov_1998}) is defined as:
\begin{align}\label{eqa17.0.1}
    S_\text{Hol} =& 2\int_\mathcal{M} d^4x \left( P^i_a \partial_t A^a_i+A^a_0 \mathcal{G}_a+N^i \mathcal{H}_i+\undertilde{N} \mathcal{H}\right)
\\
    \widetilde{P}^i_a =& i(\widetilde{E}{}^i_a-i\epsilon_a{}^{bc} \widetilde{E}{}^i_b \chi_c)
\nonumber \\
    \mathcal{G}_a =& \nabla_i \widetilde{P}^i_a = \partial_i \widetilde{P}^i_a-\epsilon_{abc} A^b_i \widetilde{P}^{ci}
\nonumber \\
    \mathcal{H}_i =& -2i\widetilde{E}{}^k_a \mathcal{F}^a_{ki}-\epsilon_{ijk}\widetilde{E}{}^i_a \widetilde{E}{}^k_b \epsilon^{lmn}\undertilde{E}{}^d_l \chi_d \mathcal{F}^{ab}_{mn}
\nonumber \\
    \mathcal{H} =& 2\widetilde{E}{}^i_a \widetilde{E}{}^j_b \mathcal{F}^{ab}_{ij}
\nonumber
\end{align}
As we can see there is an additional constraint, $\mathcal{G}$, called the Gauss constraint, which enforces the tetrad postulate. To get rid of the unwanted parameter $\chi$ in the action, one poses the change of variable $\mathcal{N}^i_D = N^i+\frac{1}{1-\chi^2} \widetilde{E}{}^i_a \chi^a (N^j \undertilde{E}{}^b_j \chi_b-\undertilde{N})$ and $\undertilde{\mathcal{N}}= \frac{1}{1-\chi^2}(\undertilde{N}-N^i \undertilde{E}{}^a_i \chi_a)$. One finds the Ashtekar action:
\begin{align}\label{eqa18.0.1}
    S_\text{A} =& 2\int_\mathcal{M} d^4x \left( P^i_a \partial_t A^a_i+A^a_0 \mathcal{G}_a+\mathcal{N}_D^i H_i+\undertilde{\mathcal{N}} H\right)
\\
    H_i =& -2\widetilde{P}^k_a \mathcal{F}^a_{ki}
\nonumber \\
    H =& 2\widetilde{P}^i_a \widetilde{P}^j_b \mathcal{F}^{ab}_{ij}
\nonumber
\end{align}
Crucially for the path integral formalism of the quantum theory, the determinant of this change of variable is perfectly compensated by the change of variable $E^i_a \leadsto P^i_a$. The Palatini action is just the real part of the Ashtekar action. In principle, one should impose reality constraints on the variables. These constraints ensure that the metric is real and its evolution with time always gives a real metric. We would like to be able to have the Palatini action as a limit of the Holst action, and not just as a real part, so that it is easier to recover the former from the latter. To do so, we switch to the Holst action with real Barbero-Immirzi parameter $\beta$, so that we now have the Palatini action with a $\mathbb{R}^2$-valued gauge field $A^X_i$:
\begin{equation}\label{eqa19.0.1}
    S_\text{Pal} = \lim_{\beta \rightarrow \infty} S^\beta_\text{Hol} = \lim_{\beta \rightarrow \infty} \left( \Re(S_\text{A})-\frac{1}{\beta}\Im(S_\text{A}) \right)
\end{equation}
To have a real formalism, one uses pairs of variables instead of complex numbers and extends the Latin indices of the beginning of the alphabet like so: $a \leadsto X\equiv (X,4)$ with $X=(a,a')$. For example $A_i^{X4} = \sqrt{\frac{\Lambda}{3}}P^X_i$, or $P_{4}^i = 0$. The Holst action is expressed as (we specify the fields right after):
\begin{align}\label{eqa20.0.1}
    S_\text{Hol}^\beta =& \int_\mathcal{M} d^4x \left( \widetilde{P}_{(\beta)}{}^i_X \partial_t A^X_i+\mathcal{N}_G^X \mathcal{G}_X+\mathcal{N}^i_D H_i +\undertilde{\mathcal{N}}H \right)
\\
    \mathcal{G}_X =& \partial_i \widetilde{P}_{(\beta)}{}^i_X + f^Z{}_{XY}A^Y_i \widetilde{P}_{(\beta)}{}^i_Z
\nonumber \\
    H_i =& -\widetilde{P}_{(\beta)}{}^j_X F^X_{ij}
\nonumber \\
    H =& -\frac{1}{2} \widetilde{P}_{(\beta)}{}^i_X \widetilde{P}_{(\beta)}{}^j_Y f_Z{}^{XY} F^Z_{ij}
\nonumber \\
    F^X_{ij} =& \partial_i A^X_j -\partial_j A^X_i+f^X{}_{YZ}A^Y_i A^Z_j
\nonumber
\end{align}
This action is the one we will use at the beginning of the quantization, and take the limit $\beta \rightarrow \infty$ to proceed further. The structure constants $f^{XYZ}$ are given in \cite{Alexandrov_2000}, and we do not specify them explicitly because one can proceed formally using only their properties. Their indices are lowered or raised with the Killing form $g_{XY} \equiv \text{diag}(\delta_{ab},-\delta_{ab})$. Moreover, the fields appearing in (\ref{eqa20.0.1}) are:
\begin{align}\label{eqa21-22-23-24-25.0.1}
    \mathcal{G}_X =& (\Im(\mathcal{G}^{(\text{A})}_a) , \Re(\mathcal{G}^{(\text{A})}_a))
\\
    A^X_i =& (\zeta^a_i,\xi^a_i)
\\
    \widetilde{P}{}^i_X =& (\widetilde{E}{}^i_a,\epsilon_{a}{}^{bc}\widetilde{E}{}^i_b \chi_c)
\\
    \widetilde{Q}{}^i_X =& (-\epsilon_{a}{}^{bc}\widetilde{E}{}^i_b \chi_c , \widetilde{E}{}^i_a)
\\
    \widetilde{P}_{(\beta)}{}^i_X =& \widetilde{P}{}^i_X - \frac{1}{\beta} \widetilde{Q}{}^i_X
\end{align}
Then, one introduces the projectors $\widetilde{P}{}^i_X = \Pi^Y_X \widetilde{Q}{}^i_Y$ and $\widetilde{P}_{(\beta)}{}^i_X = \Lambda^Y_X \widetilde{Q}{}^i_Y$, and from them one can derive the second-class constraints:
\begin{align}\label{eqa26-27.0.1}
    \phi^{ij} =& \Pi^{XY} \widetilde{Q}{}^i_X \widetilde{Q}{}^j_Y
\\  
    \psi^{ij} =&  2f^{XYZ}\widetilde{Q}{}^n_X \left[\widetilde{Q}{}^j_Y \partial_n \widetilde{Q}{}^i_Z + \widetilde{Q}{}^i_Y \partial_n \widetilde{Q}{}^j_Z \right] -\left( g^{XW}\delta^Y_U-g^{YW}\delta^X_U \right)\widetilde{Q}{}^n_X \left[\widetilde{Q}{}^j_Y  A^U_n \widetilde{Q}{}^i_W + \widetilde{Q}{}^i_Y A^U_n \widetilde{Q}{}^j_W \right]
\end{align}
The constraint $\phi^{ij}$ could be interpreted as the orthogonality of $\widetilde{Q}$ and $\widetilde{P}$ if only the matric $\Pi^{XY}$ was anti-symmetric. However, this is not the case because of the killing form $g_{XY}= \text{diag}(\delta_{ab},-\delta_{ab})$. In fact, solving $\phi^{ij}=0$ amounts to set $\chi=0$ everywhere. The quantity $\psi^{ij}$ is a constraint on the expression of $A^X_i$. Upon solving the second constraint, we introduce the metric $G^{ijkl} = 2\gamma^{ij}\gamma^{kl}-\gamma^{ik}\gamma^{jl}-\gamma^{il}\gamma^{jk}$, and after a straightforward implementation of $\phi^{ij}=0$ into $\psi^{ij}$, we find that we can add a term $K$ of $A$ provided it satisfies $\Pi^Z_U \widetilde{P}^k_Z K^U_n G^{ijn}{}_k =0$. In the process of quantization via Feynman path integrals, implementing the second-class constraints essentially amounts to inserting Dirac deltas of the constraints and the Pfaffian of their Poisson brackets into the path integral. For our practical purpose, the Pfaffian is equal to the square root of the determinant of its argument. So we are searching for $\Delta = \{\theta^{A},\theta^{B}\}_\text{P.B.}$, where $\theta^A = (\phi^{ij},\psi^{ij})$. Fortunately, \cite{Alexandrov_2000} gives it to be of the form:
\begin{equation}\label{eqa29.0.1}
    \Delta = \left( \begin{matrix}
        0 & D_1 \\
        -D_1 & D_2
    \end{matrix} \right)
\end{equation}
For $(D^{-1}_1)_{ij,kl} = \frac{1}{8} \left( 1+ \frac{1}{\beta^2} \right)[(\undertilde{Q}\undertilde{Q})_{ij}(\undertilde{Q}\undertilde{Q})_{kl}-(\undertilde{Q}\undertilde{Q})_{ik}(\undertilde{Q}\undertilde{Q})_{jl}-(\undertilde{Q}\undertilde{Q})_{il}(\undertilde{Q}\undertilde{Q})_{jk}]$, and $D_2$ is unimportant for the following because the determinant of $\Delta$ is $\det(\Delta) = \det(D_1^{ijpq}D_{1,pq}{}^{kl})$. We gave $D^{-1}$ rather than $D$ because its form resembles an inverse DeWitt supermetric. So the product is straightforward to calculate and gives $(D^{-1}_1)_{ij,kl}(D^{-1}_1)^{kl}{}_{pq} = \frac{1}{8} \left( 1+ \frac{1}{\beta^2} \right)(D_1^{-1})_{ij,pq}$. Consequently, the determinant we were searching for is $\det(\Delta) = \left\{\frac{64}{(1+\beta^{-2})^2}\right\}^3 \det(\gamma_{ij}\gamma_{kl}-\gamma_{ik}\gamma_{jl}-\gamma_{il}\gamma_{jk})^{-1}$. To obtain something entirely analogous to (\ref{eqa10}) but with a cosmological constant, we must get rid of $\chi$ and $\zeta$ (indeed, by expanding the Gauss constraint we obtain that $\zeta$, and not $\xi$, should disappear), and this is due to the first constraint $\phi^{ij}$. The second constraint adds $K$ to $A$ and we find for the Palatini Hamiltonian constraint:
\begin{align}\label{eqa30.0.1}
    H =&  - \widetilde{E}^i_a \widetilde{E}^j_b \left(R^{ab}_{ij}(\Omega)-\frac{\Lambda}{3} (E^a_i E^b_j - E^a_j E^b_i )\right)-(\widetilde{E}^i_a \widetilde{E}^j_b - \widetilde{E}^j_a \widetilde{E}^i_b) K^{a}_i K_{j}^b
\end{align}
Where we omitted the term depending on the Barbero-Immirzi parameter because we will take this parameter to be arbitrarily large in the quantization section. Moreover, we defined $\Omega\equiv A-K$. To conclude this section, notice that $\partial_t \widetilde{P}^i_X A^X_i = \partial_t \widetilde{E}^i_a K^a_i$ when solving the second class constraints.

\subsection*{3) Worldline formalism}

The worldline formalism is based on the observation that the integral along the Schwinger parameter $t$ of the heat kernel of the operator $\partial_t-\square+m^2$ is the propagator of a scalar field of mass $m$ in Quantum Field Theory. For the simple elliptic differential operator $\square$, we can write the heat kernel $K(t;x',x)=e^{t \square}\delta(x'-x)$ as follows:
\begin{equation}\label{eqa19}
    K(t;x',x)=\frac{e^{-\frac{1}{4\alpha}\frac{|x'-x|^2}{t}}}{(4\pi t)^2} = \int_{x(0)=x}^{x(t)=x'}\mathcal{D}x\,e^{-S[x^\mu]}
\end{equation}
With $S[x^\mu]=\int_0^t \frac{\dot{x}^2}{4\alpha}$ the action of a massless point particle in $\mathbb{R}^{1,3}$. The path integral has been normalized to obtain the $(4\pi t)^{-2}$ factor. In this view, the heat kernel is the Weierstrass transform of a Dirac delta, which spreads over time. Worldline formalism is a powerful tool for obtaining the generating function of the connected diagrams of QFTs. For example, the scalar QED in the Euclidean signature has for such a functional of one loop connected diagrams $\Gamma[A] = \text{Tr} [\ln(-|\partial_\mu+ieA_\mu|^2+m^2)]$. We can express this as:
\begin{align*}\label{eqa20}
    \Gamma[A] =& \text{Tr} [\ln(-|\partial_\mu+ieA_\mu|^2+m^2)]
\nonumber \\
    =& \int_0^T dT\frac{1}{T}\text{Tr}\left[ e^{-T [-|\partial_\mu+ieA_\mu|^2+m^2]} \right]
\end{align*}
\begin{align}
    =& \int_0^T dT\frac{1}{T}\int dx\, \langle x| e^{-T [-|\partial_\mu+ieA_\mu|^2+m^2]} | x \rangle
\nonumber \\
    =& \int_0^T dT\frac{1}{T}\oint \mathcal{D}x\,e^{-\int_0^T dt \left(\frac{1}{4T}\delta_{\mu \nu}\dot{x}{}^\mu \dot{x}{}^\nu+ie\dot{x}{}^\mu A_\mu(x(t))-Tm^2 \right)} 
\end{align}
It is now crystal clear that the diagrams are loops punctured with external contributions of the gauge field because the path integral has periodic boundary conditions. The counterpart of this action for a point particle in curved space is \cite{edwards2019quantum,corradini2021spinning}:
\begin{equation}\label{eqa21}
    S[e,x,g] = \int_0^1 dt \frac{1}{2}\left( \frac{1}{e}g_{\mu \nu}\dot{x}{}^\mu \dot{x}{}^\nu+e(m^2+\lambda R) \right)
\end{equation}
With $\lambda R$ a non-minimal coupling of the scalar field. The quantity $e$ is called the ``einbein'' and usually, one fixes it such that $\int_0^1 dt e=2T$. However, there is a divergence problem if we use the measure $\mathcal{D}x$. To avoid such problems, one usually uses the measure $\prod_t d x(t)\sqrt{g(x(t))}$. To make the total path integral resemble the familiar form of a Gaussian integral, we use Lee-Yang ghosts:
\begin{equation}\label{eqa22}
    \prod_t \sqrt{g(x(t))} = \int\mathcal{D}\mathfrak{a}\mathcal{D}\mathfrak{b}\mathcal{D}\mathfrak{c}\,e^{-\frac{1}{4T}\int_0^1dt\,g_{\mu \nu}(\mathfrak{a}^\mu \mathfrak{a}^\nu+\mathfrak{b}^\mu \mathfrak{c}^\nu)}
\end{equation}
where $\mathfrak{a}$ is bosonic and $\mathfrak{b}$, $\mathfrak{c}$ are fermionic. Action (\ref{eqa21}) has a corresponding Hamiltonian action of the following form:
\begin{equation}\label{eqa23}
    S[e,x,g] = \int_0^1 dt \left( p_{\mu}\dot{x}{}^\mu +\frac{e}{2}(g^{\mu \nu}p_\mu p_\nu+m^2+\lambda R) \right)
\end{equation}
Note a striking resemblance to (\ref{eqa13}). The einbein plays the role of a Lagrange multiplier and its variation gives the ``Klein-Gordon equation in momentum space'' with a non-minimal coupling (The quotation marks are here to signify that the Ricci scalar and the metric have not been projected in momentum space, so this is not the true Klein-Gordon equation in momentum space). In the position space, the metric depends on the position $x^\mu(t)$. Thus, the path integral is usually performed perturbatively by expanding the position $x^\mu(t)=\overline{x}{}^\mu+y^\mu(t)$ and then Taylor expanding the metric.\\
\indent The quantum field theoretic propagator has the same form as (\ref{eqa20}), except that the integration over $T$ does not have the $\frac{1}{T}$ factor in front of the exponential. For a theory with a potential, such as the scalar QED example above, one must modify (\ref{eqa19}) and use an expansion in terms of the Seeley-DeWitt (or HaMiDeW) coefficients:
\begin{equation}\label{eqa24}
    K_V(t;x',x)=\frac{e^{-\frac{1}{4\alpha}\frac{|x'-x|^2}{t}}}{(4\pi t)^2} \sum_{n=0}^\infty a_n(V)t^n = \int_{x(0)=x}^{x(t)=x'}\mathcal{D}x\,e^{-S[x^\mu, V]}
\end{equation}
In curved space, it has a more complicated form involving the Synge world function as a geodesic distance, and the Van Vleck-Morette determinant \cite{Vassilevich_2003}. Interestingly, in the Lorentzian signature, the heat kernel of the Laplacian $\Delta$ is the propagator of the free theory in the sense of regular Quantum Mechanics.

\section*{II. Superspace Worldline formalism and Holography}

\subsection*{1) Heat Kernel}

We already noted in a previous subsection the similarities between (\ref{eqa23}) and (\ref{eqa13}). Indeed, ignoring the momentum constraint in the gravity case, the Hamiltonian constraint has the same form as the constraint of a point particle: $g^{\mu \nu}p_\mu p_\nu+m^2 \longleftrightarrow G_{ijkl}\pi^{ij}\pi^{kl}-\sqrt{|\gamma|}\,{}^{(3)}R$. In this correspondence, $-\sqrt{\gamma}\,{}^{(3)}R$ can be seen as local inertia of the spatial metric $\gamma$. Thus, we can interpret the lapse function as a sort of einbein. We would like to further identify the phase space path integral of the quantum Einstein Gravity as a worldline formalism in the superspace (Here and throughout this article the word ``superspace'' refers to a space of metrics with a modulo on diffeomorphisms). We recall that the Holst action is expressed as
\begin{equation}\label{eqa26}
    S_\text{Hol}^\beta = \int_\mathcal{M} d^4x \left( \widetilde{P}_{(\beta)}{}^i_X \partial_t A^X_i+\mathcal{N}_G^X \mathcal{G}_X+\mathcal{N}^i_D H_i +\undertilde{\mathcal{N}}H \right)
\end{equation}
In the following and until the end of this work, we use a partition function with fixed boundary conditions. Thus, the boundary term is not integrated over and provides only a classical contribution. As we will see later, only the counter-term contributes to the entropy of the boundary if we adopt a particular method, in contrast to the usual calculation in Euclidean Quantum Gravity. The partition function with fixed boundary conditions (B.C.) of the theory is (we specify $S_\text{ghost}$ afterward):
\begin{equation}\label{eqa27}
    Z[\text{B.C.}]=\int_\text{B.C.} \mathcal{D} A^X_i \mathcal{D} \widetilde{P}_{(\beta)}{}^i_X \mathcal{D} \mathcal{N}^\alpha \mathcal{D}\overline{c}_\alpha \mathcal{D} c{}^\alpha\,\sqrt{\det(\Delta)} \delta(\psi^{ij})\delta(\phi^{ij})\delta(f^\alpha+g^\alpha)\,e^{iS^\beta_\text{Hol}+iS_\partial+iS_\text{ghost}[\overline{c}_\alpha,c^\alpha]}
\end{equation}
To find $S_\text{ghost}$, one uses the BRST formalism. The BRST formalism is a general setting for gauge theories in which one introduces ghostly parameters, and can be viewed as a generalization of the Faddeev-Popov method. Indeed, due to gauge freedom, the naive path integral diverges, and to cure this property, one has to integrate over the space of fields modulo the gauge freedom. But since it is hard to find in practice, we integrate over the whole space of fields and correct the integrand of the path integral to have the same result as if we had integrated over gauge inequivalent configurations only. The purpose of our work is not to thoroughly explain the BRST setting used, as it simply is a tool to achieve our goals, and not these goals \textit{per se}. We mainly reuse the notations and the method of \cite{Alexandrov_1998}. Let $\Phi_\alpha$ be our constraints (Gauss, spatial diffeomorphisms, and Hamiltonian), and $n^\alpha$ their associated Lagrange multipliers. Then one extends the phase space by introducing $\pi$ and $(\overline{b},c,b,\overline{c})$ which are the canonical momentum of the Lagrange multipliers, and the ghosts, respectively. The ghosts $\overline{b}$, $b$ are complex Grassmann numbers while $\overline{c}$, $c$ are real Grassmann numbers. Thus, we have the Lagrangian density:
\begin{equation}\label{eqa28}
    L_\text{eff}= \dot{q}{}^s p_s + \dot{n}{}^\alpha \pi_\alpha+\dot{c}{}^\alpha \overline{b}_\alpha+\dot{b}{}^\alpha \overline{c}_\alpha-H_\text{eff}
\end{equation}
Where $q$ and $p$ are our canonical conjugated variables of the original phase space. One has $H_\text{eff} = -\{\psi,\Omega\}_+$, where $\psi$ is called the gauge fixing fermion (not to be confused with the second class constraint $\psi^{ij}$) and $\Omega$ the BRST generator. The power of BRST formalism resides in its gauge fixing fermion which we are free to choose, provided that it is indeed of ghost number -1. We use the same gauge fixing fermion as \cite{Alexandrov_1998,Alexandrov_2000} to have:
\begin{align}\label{eqa28-29.1}
    \Omega =& -ib^\alpha \pi_\alpha+ c^\alpha \Phi_\alpha+\frac{1}{2}c^\alpha c^\beta C^\gamma_{\alpha \beta} \overline{b}_\gamma+c^\alpha c^\beta c^\gamma U^{(2)\delta \lambda}_{\alpha \beta \gamma} \overline{b}_\delta \overline{b}_\lambda
\\
    \psi =& -\overline{b}_\alpha n^\alpha+i\overline{c}_\alpha\left( \frac{1}{\gamma}f^\alpha(q,p)+\frac{1}{\gamma} g^\alpha(n)\right)
\end{align}
The coefficients $C^{\gamma}_{\alpha \beta}$ are the Poisson brackets between the constraints, and in our case of gravity, these are problematic when $\alpha,\beta = H$, (that is to say when we take the Poisson bracket of the Hamiltonian constraint with itself). $U^{(2)}$ is similarly found upon nesting the Poisson brackets of the constraints $\Phi$. There are no higher-order terms because all the other coefficients are identically zero. In the gauge fixing fermion, two sets of functions have been introduced: $f$ and $g$ which depend on the canonical variables of the original phase space, and the Lagrange multipliers, respectively. Thus, one finds:
\begin{align}\label{eqa28.2}
    H_\text{eff} =& -n^\alpha \Phi_\alpha-i\overline{b}_\alpha b^\alpha+c^\alpha n^\beta C^\gamma_{\alpha \beta} \overline{b}_\gamma -3 c^\alpha c^\beta n^\gamma U^{(2) \delta \lambda}_{\alpha \beta \gamma} \overline{b}_\delta \overline{b}_\lambda+ \frac{1}{\gamma} \left[ (f^\alpha+g^\alpha)\pi_\alpha-\overline{c}_\alpha \frac{\partial g^\alpha}{\partial n^\beta} b^\beta-i\overline{c}_\alpha \{f^\alpha,\Phi_\beta\} c^\beta \right.
\nonumber \\    
    &\left.-i\overline{c}_\alpha \{f^\alpha,C^\delta_{\beta \gamma} \} c^\beta c^\gamma \overline{b}_\delta-i\overline{c}_\alpha \{f^\alpha,U^{(2)\xi \eta}_{\beta \gamma \delta}\}c^\beta c^\gamma c^\delta \overline{b}_\xi \overline{b}_\eta \vphantom{\frac{\partial}{\partial}}\right]
\end{align}
Upon using a change of variable $\pi \leadsto \gamma \pi$, and $\overline{c} \leadsto \gamma \overline{c}$, and the taking the limit $\gamma \rightarrow 0$, one can easily integrate over $\pi$, $b$ and $\overline{b}$. To further simplify $L_\text{eff}$, we impose what is called in \cite{Alexandrov_1998} a ``Yang-Mills gauge'', that is, a choice of $f$ and $g$ so that the ghost action has no term higher than $\overline{c}_\alpha [\cdots] c^\alpha$. We choose $f(q,p)=f=c^\text{ste}$, and $g^\alpha(n) = (-f^1,-f^2,-\mathcal{N}_D^i,-\undertilde{\mathcal{N}})$ so that the condition to have a Yang-Mills gauge is fulfilled. Thus, we have:
\begin{align}\label{eqa28.3}
    L_\text{eff} \leadsto& \dot{q}{}^s p_s+ n^\alpha \Phi_\alpha-\underbrace{i\overline{c}_\mu \left( \delta^\mu_\alpha\partial_t- C^\mu_{\alpha \lambda}n^\lambda \right)c^\alpha}_{L_\text{ghost}}
\end{align}
Here, $\mu$ and $\alpha$ are multi-indices encapsulating the indices of their respective constraints. For example, $g^{\mu=H} = -\undertilde{\mathcal{N}}$, and $g^{\mu=D_i} = -\mathcal{N}^i_D$. To expand further the ghost action, we have to know which of the structure functions $C^\lambda_{\alpha \beta}$ are zero. Due to our choice of $f$ and $g$, we have only three non-zero structure functions, given by:
\begin{align}\label{eqa28.4-5-6}
    \{\Phi_{D}[\vec{N}],\Phi_{D}[\vec{M}]\}_\text{P.B.} =& \Phi_{D}[\pounds_{\vec{N}}\vec{M}]
\\
    \{\Phi_{D}[\vec{N}],\Phi_{H}[M]\}_\text{P.B.} =& \Phi_{H}[\pounds_{\vec{N}}M]
\\
    \{\Phi_{H}[N],\Phi_{H}[M]\}_\text{P.B.} =& \Phi_{D}[N\overleftrightarrow{\partial}^i M]
\end{align}
Where the constraints are smeared with their corresponding Lagrange multipliers. Thus, we obtain:
\begin{align}\label{eqa28.7}
    S_\text{ghost} =& -i\int_\mathcal{M} d^4x\, \overline{c}_{D_i} \left( \delta^{D_i}_{D_k}\partial_t -C^{D_i}_{D_k D_j}\mathcal{N}_D^j \right)c^{D_k}+i\int_\mathcal{M} d^4x\, \overline{c}_{D_i} C^{D_i}_{H H}\undertilde{\mathcal{N}} c^{H}
\nonumber \\
    & -i\int_\mathcal{M} d^4x\, \overline{c}_{H} \left(\partial_t -C^{H}_{H D_j}\mathcal{N}_D^j \right)c^{H} +i\int_\mathcal{M} d^4x\, \overline{c}_{H} C^{H}_{D_j H} \undertilde{\mathcal{N}} c^{D_j}
\end{align}
According to the end of I. 2), and upon solving the constraints $\phi^{ij}$ and $\psi^{ij}$, and posing $\beta \rightarrow \infty$ (as pointed out in \cite{Alexandrov_2000}, the different factors arising from $\beta$ in the path integral perfectly cancel out), we can rewrite (\ref{eqa27}) and find:
\begin{align}\label{eqa28.8}
    Z[\text{B.C.}] =& \int_\text{B.C.} \mathcal{D} K^a_i \mathcal{D} \widetilde{E}^i_a \mathcal{D} \mathcal{N}^\alpha \mathcal{D}\overline{c}_\alpha \mathcal{D} c{}^\alpha\,\sqrt{\det(\Delta)} \delta(f^\alpha+g^\alpha)\,e^{iS_\text{Pal}+iS_\partial+iS_\text{ghost}}
\\
    S_\text{Pal} =& \int_\mathcal{M} d^4x \left( -K^a_i \partial_t\widetilde{E}^i_a +\mathcal{N}_G^a \mathcal{G}_a+\mathcal{N}^i_D H_i +\undertilde{\mathcal{N}}H \right)
\nonumber \\
    \mathcal{G}_a =& \partial_i \widetilde{E}^i_a + \epsilon^c{}_{ab}\Omega^b_i \widetilde{E}^i_c,\,\,\, \Omega = A-K
\nonumber \\
    H_i =& -\widetilde{E}^j_c \epsilon^c{}_{ab} R^{ab}_{ij}(\Omega)
\nonumber \\
    H =&  - \widetilde{E}^i_a \widetilde{E}^j_b \left(R^{ab}_{ij}(\Omega)-\frac{\Lambda}{3} (E^a_i E^b_j - E^a_j E^b_i)\right)-(\widetilde{E}^i_a \widetilde{E}^j_b - \widetilde{E}^j_a \widetilde{E}^i_b) K^{a}_i K_{j}^b
\nonumber
\end{align}
Integrating on $\mathcal{N}^\alpha$ and $K^a_i$ we obtain the partition function:
\begin{align}\label{eqa28.10}
    Z[\text{B.C.}] =& \int_\text{B.C.} \mathcal{D} E^i_a \mathcal{D} \mathcal{N}^\alpha \mathcal{D}\overline{c}_\alpha \mathcal{D} c{}^\alpha\,\frac{1}{\sqrt{\det(\undertilde{\mathcal{N}}\delta^i_a \delta^j_b)}}\delta(f^\alpha+g^\alpha) \,e^{iS'_\text{Pal}[\mathcal{N},E]+iS_\partial[\mathcal{N},E]+iS_\text{ghost}[\mathcal{N},\overline{c},c]}
\nonumber \\
    =& \delta(0)\delta(0)\int_\text{B.C.} \mathcal{D} E^i_a \mathcal{D}\overline{c}_\alpha \mathcal{D} c{}^\alpha\,\frac{1}{\sqrt{\det(f^H\delta^i_a \delta^j_b)}}\,e^{iS'_\text{Pal}[f,E]+iS_\partial[f,E]+iS_\text{ghost}[f,\overline{c},c]}
\end{align}
With $S'_\text{Pal} = -\int_\mathcal{M} d^4x\left[ \frac{\kappa}{2\mathcal{N}}G^{ab}_{ij}\partial_t \widetilde{E}{}^i_a \partial_t \widetilde{E}{}^j_b + \frac{1}{2\kappa}\mathcal{N}\sqrt{|\gamma|}\left( {}^{(3)}R-2\Lambda \right) + \text{cnstr}\right]$, and ``cnstr'' is the diffeomorphism constraint. Notice that we integrate on $E^i_a$ and not on $\widetilde{E}{}^i_a$, because the Pfaffian of the second class constraints and the integral on $K^a_i$ do not compensate exactly. In fact, there is a remaining factor $\det(\sqrt{\gamma})^{-1}$, which is precisely the inverse of the jacobian of the transformation $E^i_a \leadsto \widetilde{E}{}^i_a$. Due to our choice of $f$ and $g$ in the gauge fixing fermion, we have two Dirac deltas evaluated at $0$. This is not that problematic because, in the end, we are interested in ratios of partition functions. Alternatively, we could have posed $g^\alpha(n)=(0,0,-\mathcal{N}_D^i,-\undertilde{\mathcal{N}})$, and integrate over $f^1$ and $f^2$ to get rid of these parameters. Now, we rescale $f^H \leadsto \frac{1}{\lambda} f^H$, $t \leadsto \lambda t$, $c^H \leadsto \frac{1}{\lambda} c^H$ and $\overline{c}_H \leadsto \lambda\overline{c}_H$, for $\lambda \rightarrow 0$. Notice that the Hamiltonian constraint is invariant by this change, but not the diffeomorphism constraint which vanishes in the limit $\lambda \rightarrow 0$. Expressing $\sqrt{\det(f^H\delta^i_a \delta^j_a)}^{-1}$ as the integral of a Gaussian, we find that this term is invariant too. The structure functions in (\ref{eqa28.7}) are not invariant in general, but change only for $C^{D_i}_{HH} \leadsto \lambda^2C^{D_i}_{HH}$. Thus, in the limit $\lambda \rightarrow 0$ we obtain:
\begin{align}\label{eqa28.11}
    S_\text{ghost} \stackrel{\lambda \rightarrow 0}{=}& -i\int_\mathcal{M} d^4x\, \overline{c}_{D_i} \delta^{D_i}_{D_k}\partial_t c^{D_k} - i\int_\mathcal{M} d^4x\, \overline{c}_{H} \partial_t c^{H}
\end{align}
So integrating the remaining BRST ghosts simply gives a determinant of $\partial_t$. Now, we introduce boundary conditions such that $E(0)=E_0$, $E(\beta_m) = E_{\beta_m}$ and $f^H(x\in I\times \partial \Sigma)=0$, with $\beta_m \ll 1$. Moreover, we pose $E_{\beta_m} = E_0+\Delta E \frac{t}{\beta_m}$, in the spirit of the Worldline formalism. Indeed, upon noticing that the Hamiltonian constraint is quadratic in the time derivative of the ``position'' variable $E^i_a$ and that the partition function resembles an integral of a heat kernel in the space of $E^i_a$, we identify the following expression as the result of a worldline formalism in this space.
\begin{align}\label{eqa28.12}
    Z[\text{B.C.}]|_0^{\beta_m} \propto&\frac{e^{iS_\partial[E,f]}}{\sqrt{\det_\Sigma(\beta_mf^H\delta^i_a \delta^j_b)}}\,e^{-i\beta_m\int_\Sigma d^3x\left[ \frac{\kappa}{2\beta^2_m f^H}(G_0+\Delta G)^{ab}_{ij}\Delta\widetilde{E}{}^i_a \Delta \widetilde{E}{}^j_b + \frac{1}{2\kappa}f^H\sqrt{|\gamma_0|}\left( {}^{(3)}R_0-2\Lambda +\Delta{}^{(3)}R(\beta_m)\right)\right]}
\end{align}
With $\Delta{}^{(3)}R(\beta_m)$ and $\Delta G$ the remaining parts of the expansion of the 3D Einstein-Hilbert action and the metric $G^{ab}_{ij}$ about $E_0$, respectively. But this partition function is only for a short interval $\beta_m \ll1$. The idea is to get rid of the $\Delta E$ terms and integrate over $E_0$ to have the partition function of 3D gravity, which may be simpler to deal with. To achieve this goal, we use a closed-time contour. Since we are in Lorentzian signature, we choose a Keldysh contour. \\
\indent We write by $\mathcal{T}_c$ the time ordering along the closed time contour starting from $t=n\beta_m$. Let us denote by $\oint_{c_n}$ the small closed contour integration, going from $t=n\beta_m$ to $t=(n+1)\beta_m$ and then coming back to $t=n\beta_m$. We can decompose formally the operator $\mathcal{T}_c e^{i\oint_{c_0}dt \mathcal{H}}$ as:
\begin{equation}\label{eqa31}
    \left.\mathcal{T}_ce^{i\oint_{c_0}dt\mathcal{H}} \right|_\text{B.C.} = (\delta^c(0))^2\frac{\det_c(\partial_t)\, e^{iS_\partial[E,f]}}{\sqrt{\det_\Sigma(\beta_m f^H\delta^i_a \delta^j_b)}} \int_\text{B.C.} \mathcal{D}\mu\,e^{-2i\beta_m\int_\Sigma d^3x \frac{1}{2\kappa}f^H\sqrt{|\gamma_0|}\left( {}^{(3)}R_0-2\Lambda \right)}| \Psi_{[R]} \rangle \langle \Psi_{[R]} |
\end{equation}
With $\mathcal{D} \mu$ a formal integration measure that is used to integrate over gauge inequivalent configurations. Since the heat kernel (\ref{eqa28.12}) with closed time contour can be thought of as the trace of an evolution operator, we write $Z[\text{B.C.}]|_{c_0} \propto \text{tr}\left[ \left.\mathcal{T}_ce^{i\oint_{c_0} dt\mathcal{H}}\right|_\text{B.C.} \right]$, with $\mathcal{H}$ a Hamiltonian quadratic in $\frac{\delta}{\delta E}$, and $\text{tr}$ a trace operation we define to be over states satisfying all the first-class constraints of $S_\text{Pal}$. These states are given by the Kodama state \cite{Kodama} when working with the Ashtekar variables. The problem is, that the Kodama state is just one state and is non-normalizable \cite{witten2003note}, among other problems. However, for the Holst theory, there exist generalizations of the Kodama state \cite{randono2006_1, randono2006_2}. This generalization is normalizable and solves or invalidates the other problems. According to \cite{randono2006_2}, the physical inner product between two states separated by an interval $[0,\beta_m]$ is ($[R]$ is the equivalence class modulo spatial diffeomorphisms of configurations of space containing the configuration $R$):
\begin{equation}\label{eqa31.1}
    \langle \Psi_{[R']_{\beta_m}}| \Psi_{[R]_0} \rangle_\text{phys} = e^{-i\frac{3}{4\kappa \Lambda}S_\text{Top}} Z[\text{B.C.}]|_0^{\beta_m}
\end{equation}
Indeed, \cite{randono2006_2} uses the MacDowell-Mansouri action to generate a WBK approximation to the state, which turns out to also be an exact solution to the constraints. But, the MacDowell-Mansouri action is $-\frac{3}{4\kappa \Lambda}S_\text{MM} = S_\text{Pal}-\frac{3}{4\kappa \Lambda}S_\text{Top}$, with $S_\text{Top} = \int_\mathcal{M} \star R \wedge R = \int_{\partial \mathcal{M}} \epsilon_{abcd}\left(\omega^{ab}\wedge d\omega^{cd}+\frac{2}{3}\omega^{ab}\wedge \omega^c{}_e \wedge \omega^{ed}\right)$. Because we integrate over constant boundaries with cyclic time, we pulled the $S_\text{Top}$ contribution out of the partition function in (\ref{eqa31.1}). Notice that if $\omega|_{\mathbb{S}^1\times \partial \Sigma} =0$, then this contribution can be zero. The references \cite{randono2006_1, randono2006_2} work with the Holst action with real, finite Barbero-Immirzi parameter $\beta$. But there seems to be no obstruction to set $\beta \rightarrow \infty$ in this work, \textit{after} having done the whole calculation. One can see this procedure as calculating (\ref{eqa31}) for the Holst theory (seen as a deformation of the Palatini theory), and then taking the limiting case of $\beta^{-1} \rightarrow 0$.\\
\indent In the limit $\beta_m \rightarrow 0$, we obtain the kinematical inner product out of the physical one:
\begin{equation}\label{eqa31.2}
    \langle \Psi_{[R']_{\beta_m}} | \Psi_{[R]_{0}} \rangle_\text{phys} \stackrel{\beta_m \rightarrow 0}{\propto} \delta(\Delta [E]^i_a) 
\end{equation}
Now, we can glue together $N$ different closed-time amplitudes with fixed boundary conditions to obtain:
\begin{align}\label{eqa32}
    \langle \Psi_{[R]_{N}} |\prod_{k=0}^N\left.\mathcal{T}_c e^{i\oint_{c_k}dt\mathcal{H}}\right|_\text{B.C.}| \Psi_{[R]_0} \rangle \propto& \int_\text{B.C.} \mathcal{D}\mu' e^{-\frac{2i \beta_m N}{2\kappa}\int_{\Sigma_{0}} d^3x f^H\sqrt{|\gamma_{0}|}\left( {}^{(3)}R_{0}-2\Lambda \right)} \langle \Psi_{[R]_{N}} | \Psi_{[R']_{N-1}} \rangle \langle \Psi_{[R']_1} |\Psi_{[R]_{0}} \rangle
\nonumber \\
    \stackrel{\beta_m \rightarrow 0}{=}& \delta^{[E]}(0) e^{-\frac{2iN\beta_m}{2\kappa}\int_{\Sigma_{0}} d^3x f^H\sqrt{|\gamma_{0}|}\left( {}^{(3)}R_{0}-2\Lambda \right)}
\end{align}
Where the proportionality constant containing $S_\partial$ has to be reintroduced. We can conclude this section by saying that taking into account the boundary conditions, we have
\begin{align}\label{eqa34}
    Z_\text{K.}[\text{B.C.}]\equiv&\, \text{tr}_{\Psi_{[R]}}\left[\prod_{k =0}^N\mathcal{T}_c \left.e^{i\oint_{c_k}dt\mathcal{H}}\right|_{\text{B.C.}}\right] \propto e^{iS_\partial[f,E]}\int_\text{B.C.}  \mathcal{D}\mu\, e^{-\frac{2iN \beta_m}{2\kappa}\int_{\Sigma_{0}} d^3x f^H\sqrt{|\gamma|}\left( {}^{(3)}R-2\Lambda \right)}
\end{align}
Where K. stands for ``Keldysh''. Sending it $N$ to infinity can be troublesome, but, as we will see in the next section, we can handle this by carefully defining the way it tends to infinity. Note that since $\mathcal{D} \mu$ is a measure over the space of triads and spin connections modulo spatial diffeomorphism, the diffeomorphism constraint is satisfied at each time of the procedure.

\subsection*{2) The ``time'' parameter}

In this part and until the end of this work, $\beta_m$ is now a non-zero parameter. This is a reusing of notations but the purpose of the old $\beta_m$ and the new one is the same: it is a time interval. We obtained an expression for the Keldysh partition function which depends on $f^H$, the parameter representing the lapse function. Our short-time expansion is plagued by this parameter which can be as large as we want. To solve this problem, let us first focus on the boundary contribution. Because the boundary action is on-shell, we can express it in whichever formalism we want. Specifically, it is convenient to express this in the first-order formalism. As a boundary metric, we use the Schwarzschild metric, as it is quite simple and compatible with a cosmological constant (indeed, in McVittie coordinates, one can set $t=0$, which corresponds to a Schwarzschild black hole). However, we directly face the problem that the lapse parameter $f^H$ is identically zero on the boundary (or rather, $f^H=0$ defines the boundary). Therefore, it appears that there is no boundary contribution. In fact, we must look closer at the quantities we have. Indeed, we can always consider $N$ such that $f^H N \beta_m|_{\partial \Sigma}=\alpha \beta_m$ for a given integer $\alpha$. However, we cannot attribute this constant value to $N$ because the bulk $f^H$ is non-zero, and thus, the heat kernel (\ref{eqa31}) would be constant in $E$.\\
\indent To overcome this problem, it is convenient to introduce the function of space-time coordinates $T(t,x) = \int_0^t f^H(s,x) ds$. Doing so we obtain $dT=f^Hdt$, and we introduce $\xi$ such that $T((k+1)\xi \beta_m)-T(k\xi\beta_m)=\beta_m$. We no longer specify the space dependence of $T$. Noticing that in (\ref{eqa31}) $\sqrt{|\gamma_0|}({}^{(3)}R_0-2\Lambda)$ does not depend on the $t$, we can safely change $f^H dt$ into $dT$ without thinking about a possible change of coordinate in the remaining of the integrand (the term in $\Delta E$ will anyway vanish by Keldysh contour). Now, we will make a key assumption, namely that, as before, $\beta_m \ll 1$ (as we will see in the discussion at the end of the next section, even in natural units, $\beta_m \ll 1$), and that $f^H=\overline{f}{}^H+\Delta f^H \frac{t-k\xi \beta_m}{\xi\beta_m}$ on the interval $[k\xi \beta_m,(k+1)\xi\beta_m]$. Then, we simply swap the places of the spatial integral and the hidden ``time'' integral in (\ref{eqa28.12}) because $T$ depends on the spatial coordinates:
\begin{equation}\label{eqa35}
     e^{2i \int_{k \xi \beta_m}^{(k+1) \xi \beta_m} dt \int_\Sigma d^3x \frac{f^H}{2\kappa} \sqrt{|\gamma_0|}({}^{(3)}R_0-2\Lambda)} \leadsto e^{2i \int_\Sigma d^3x \int_{T(k \xi \beta_m)}^{T((k+1) \xi \beta_m)}dT \frac{1}{2\kappa}\sqrt{|\gamma_0|}({}^{(3)}R_0-2\Lambda)}
\end{equation}
Thus, when $\xi=[\frac{1}{2}(f^H((k+1) \xi \beta_m)+f^H(k \xi \beta_m))]^{-1}$, we have $\sum_{k=0}^{\alpha-1}\int_{T(k \xi \beta_m)}^{T((k+1) \xi \beta_m)}dT = \alpha \beta_m$. This is a cyclic definition but the contributions due to $\xi$ cancel out, and this is why we did not specify that $\xi$ depends on $k$. In (\ref{eqa31}), this result traduces the passage of time by one unit at \textit{all} space points $x$,
\begin{equation}\label{eqa36}
     \langle \Psi_{[R']_{\xi \beta_m}} | \mathcal{T}e^{-i\frac{1}{2\kappa}\int_{\Sigma} d^3x \int_0^{\xi \beta_m}dt N\mathcal{H}}| \Psi_{[R]_0} \rangle = \prod_{x\in\Sigma} \langle \Psi_{[R']_{\xi(x) \beta_m}}(x) | \mathcal{T}e^{-i\frac{1}{2\kappa}d^3x\int_0^{\beta_m}dt'\mathcal{H}}| \Psi_{[R']_{0}}(x) \rangle
\end{equation}
Thus, for each $d^3x$, we choose a space-like hypersurface at a time $\xi(x)\beta_m$ so that it would correspond to a time $\beta_m$ if $f^H(x)=1$. Specifically, the purpose of $\xi$ is to compensate the contribution of the lapse function. Equation (\ref{eqa36}) is a complete paradigm change because it transforms a worldline formalism in the superspace into a continuum of worldline formalisms in the foliation of hypersurfaces. In a sense, our 3+1D theory is an effective theory emerging from a 3D one thanks to the Hamiltonian constraint. This changes nothing to the boundary terms but changes everything for the bulk terms. We can now safely calculate the former as
\begin{align}\label{eqa37}
    \lim_{\epsilon \rightarrow 0}2\int_{\partial \Sigma+\epsilon}d^2x\int_0^{N\beta_m}dt\, \mathcal{L}_\partial[\sigma] =& \lim_{\epsilon \rightarrow 0}\frac{1}{\kappa}\int_{\partial \Sigma+\epsilon} \int_0^{N\beta_m}dt f^H(e^a \wedge \omega^a-e_{(0)}^a \wedge \omega^a_{(0)})
\nonumber \\
    \stackrel{\text{Sch.}}{=}& \lim_{\epsilon \rightarrow 0}\int_{\partial \Sigma+\epsilon}d^2x\frac{N\beta_m}{\kappa} r^2\sin\theta\,f^H(4e^2_{\theta}\omega^{13}_{\varphi}-4e^2_{(0),\theta}\omega^{13}_{(0),\varphi})
\end{align}
Due to its expression, the on-shell boundary term seems to approach zero as $\epsilon \rightarrow 0$, but the counterterm does not. Its contribution is as follows:
\begin{align}\label{eqa38}
    \lim_{\epsilon \rightarrow 0}2\int_{\partial \Sigma+\epsilon}d^2x\int_0^{N\beta_m}dt\, \mathcal{L}_\partial[\sigma] \stackrel{\text{Sch.}}{=} \frac{4\beta_m}{8\pi }\int_{\partial \Sigma}d^2x\, r\sin\theta = 2\beta_m r_\text{Sch.}
\end{align}
This is for the interval $[0,\beta_m]$ on the boundary, which is equal to $[0,\beta_m/(f^H|_{\partial \Sigma})]$ from the point of view of the flat coordinates $(r,\theta,\varphi)$. This interval should be used, because it satisfies $N \rightarrow \infty$. Equation (\ref{eqa38}) does not correspond to the well-known Euclidean Quantum Gravity counterpart, $\frac{\beta}{4}r_\text{Sch.}$ \cite{cassaniBHlectures}, and for a good reason: this is a naive calculation that does not consider the bulk contribution. Although the standard procedure exhibits a vanishing bulk contribution, we are now inside a three-dimensional hypersurface and not inside of the full space-time. Therefore, in principle, the bulk contribution does not disappear. We will see at the end of the following subsection that upon identifying $2\beta_m$ as a time unit, we obtain the correct result if we consider the 3D bulk.

\subsection*{3) A change of variable}

We will explore the possibility of having a state $|A^+, A^-\rangle$ as an initial state, where $A^\pm$ are two gauge fields. Let $\omega$ be the spin connection and $e$ the triad (so $e=E$ and $\omega = \Omega$ of (\ref{eqa30.0.1})). We will no longer write the brackets to denote equivalence classes. Then, ignoring the $\delta(0)$ factors, we write:
\begin{equation}\label{eqa39}
    Z_\text{K.}[A^\pm|_{\partial}]\equiv \lim_{N\rightarrow \infty}\text{Tr}_{e_0,\omega_0}\left[\left.\mathcal{T}_c\,e^{-i\oint_c dt\mathcal{H}(N)}\right|_{(\omega\pm\sqrt{\Lambda}e)|_{\partial \Sigma}=A^\pm|_{\partial}}\right] \leadsto \lim_{N\rightarrow \infty}Z_{N\beta_m}^\text{3D}[A^\pm|_\partial]
\end{equation}
The new variables $A^+ = \omega+\sqrt{\Lambda}e$ and $A^- = \omega-\sqrt{\Lambda}e$ will prove to be useful because they convert the 3D Einstein-Hilbert action into the difference between two Chern-Simons theories \cite{Witten2007, Carlip1995, Chang}. Note that these variables resemble their Lorentzian Anti-de Sitter (AdS) counterparts. This is because we chose $(-,+,+,+)$ as 3+1D metric signature; thus, the 3D Ricci scalar has a $(+,+,+)$ signature, and thus it is a Euclidean 3D Ricci scalar. But since the Euclidean Einstein-Hilbert action is $\sim \int (R+2\Lambda^E)$, we have indeed $\Lambda^E=-\Lambda <0$, so an AdS case. Upon ignoring the counter-term $K_0$ in the 3D action $S_\partial$, we can write:
\begin{equation}\label{eqa40}
    Z_{N\beta_m}^\text{3D}[A^\pm|_\partial] = \int_{A^\pm|_\partial} \mathcal{D} A^+ \mathcal{D}A^-\,e^{iS_\text{CS}[A^+]-iS_\text{CS}[A^-]} = Z^\text{CS}[A^+|_\partial]Z^\text{CS}[A^-|_\partial]
\end{equation}
We omit the Gauss constraint, now written in terms of curvatures $d_\omega e = \frac{1}{2\sqrt{\Lambda}}(F^+-F^-)=0$, because $ Z^\text{CS}$ will turn out to automatically satisfy $\hat{F}{}^\pm Z^\text{CS}[A^\pm|_\partial]=0$. It is well-known that a 3D Chern-Simons theory is equivalent to a WZW model at the boundary. We will precisely address this statement by manipulating $Z^\text{CS}[A]$ ($A$ refers to either $A^+$ or $A^-$). First, we use a gauge transformation to make the WZW action appear. Under a change $A_z \leadsto hA_z h^{-1}+\partial_z hh^{-1}$, the CS action becomes \cite{dunne1999aspects,ELITZUR1989108}:
\begin{equation}\label{eqa40.1}
    S_\text{CS}[A] \leadsto S_\text{CS}[A]-\frac{k}{4\pi}\int_{\partial}\text{Tr}[A \wedge h^{-1}d h]-\underbrace{\frac{k}{12\pi}\int_{B}\text{Tr}[h^{-1}dh\wedge h^{-1}dh \wedge h^{-1}dh]}_{\equiv \Gamma_\text{WZ}[h]}
\end{equation}
Where $\Gamma_\text{WZ}[h]$ is the Wess-Zumino term. We now use the usual procedure by decomposing $A$ into $A_r+\tilde{A}$, where $A_r \equiv A_r dx^1$. The usual setting is a cylindrical universe with $A_0+\tilde{A}$ for $\tilde{A} \equiv (A_\rho, A_\theta)$. Assuming that our boundary is spherical, we can use stereographic projection to convert it into a complex plane in polar coordinates. Thus the variable $A_r$ plays the role of a ``Euclidean $A_0$''. With this stereographic projection, we can rewrite $\tilde{A} = (A_X,A_Y)$, and these are the variables we will use in the following. The Chern-Simons action $\int A\wedge F$ is, after our decomposition:
\begin{align}\label{eqa41}
    S_\text{CS}[A] = \frac{k}{4\pi}\int_{\Sigma \simeq \mathbb{R}^+\times\overline{\mathbb{C}}}\text{Tr}[\tilde{A}\wedge \partial_r \tilde{A}+A_r \wedge F_{X,Y}]
\end{align}
Where $F$ is the curvature of $A$ ($F=dA+\frac{1}{2}[A \stackrel{\wedge}{,} A]$). As we can see, $A_r$ is a Lagrange multiplier. To achieve our goals, we relax a condition on the boundary, namely the $A_r|_\partial$ condition (we can see our previous notation of $A^\pm|_\partial$ as implicitly being $(A^\pm_X|_\partial, A^\pm_Y|_\partial)$). By writing $h(z,\overline{z})=g^\dagger(\overline{z}) g(z)$, we obtain the following partition function:
\begin{align}\label{eqa40.2}
    Z^\text{CS}[A_z|_\partial] &= \int_{A_z|_\partial}\mathcal{D}A^h_z\mathcal{D}A^h_{\overline{z}}\,\delta(F^h_{z\overline{z}})\,e^{iS_\text{CS}[A]+\frac{ik}{2\pi}\int_{\partial}d^2z \text{Tr}[A_z|_\partial h^{-1}\partial_{\overline{z}}h]-i\Gamma_\text{WZ}[h]}
\nonumber \\
    &=\int_{A_z|_\partial}\mathcal{D}(\partial_z gg^{-1})\mathcal{D}(g^{\dagger,-1}\partial_{\overline{z}}g^\dagger)\,e^{kS_\text{WZW}[h]+\frac{ik}{2\pi}\int_{\partial}d^2z \text{Tr}[A_z|_\partial h^{-1}\partial_{\overline{z}}h]}
\end{align}
where we introduced the variables $z=X+iY$. Note that the Dirac delta is invariant under the change $A\leadsto A^h$. This is because it can be expressed as a Gaussian, and so will involve $F_{z\overline{z}}^2$, which is gauge invariant. The integral boundary can be completely dropped because the condition it enforces is satisfied. Now, we will closely follow \cite{Nair3DSchro} and express the integration measure differently. Rewriting $g=\rho U$, for $\rho$ a Hermitian matrix and $U$ a unitary one, we obtain:
\begin{equation}\label{eqa45}
    \mathcal{D}A_z\mathcal{D}A_{\overline{z}} = |\det(D^2)|\mathcal{D}U\mathcal{D}h,\,\,\,h=g^\dagger g = \rho^2
\end{equation}
The determinant of the elliptic differential operator $D^2$ is the Jacobian of this change in functional coordinates. More precisely, $D^2 = D_z D_{\overline{z}}$, where $D_z$ ($D_{\overline{z}}$) is the covariant derivative involving the field $A_z$ ($A_{\overline{z}}$). This determinant can be formally expressed using a propagator trace. Let $\ln \det(D) \equiv \ln \det(D_z)$ be an effective action. Then:
\begin{equation}\label{eqa46}
    S_\text{eff.}[g] \equiv \ln \det(D) = \text{Tr}\ln(D) \Rightarrow \frac{\delta S_\text{eff.}[g]}{\delta A^a_z(x)} \stackrel{!}{=} \text{Tr}\left[ \lim_{y\rightarrow x} D^{-1}(x,y)(-i\tau^a) \right]
\end{equation}
Because the propagator is divergent in the limit $y \rightarrow x$, one uses a regulator $\sigma$ to obtain a regulated propagator:
\begin{equation}\label{eqa47}
    D^{-1}_\text{reg.}(x,y) = \int_{\partial \Sigma} d^2z\frac{g(x)g^{-1}(z)}{\pi(\overline{x}-\overline{z})} \sigma(y,z;\epsilon),\,\,\,\sigma(y,z;\epsilon)= \frac{\varrho(y,z)}{\pi \epsilon}e^{-\frac{|y-z|^2}{\epsilon}}
\end{equation}
Where we noted that $\partial_{\overline{z}}^{-1}(x,y)= \frac{1}{\pi(\overline{x}-\overline{y})}\Rightarrow D^{-1}(x,y) = \frac{g(x)g^{-1}(y)}{\pi(\overline{x}-\overline{y})}$. Note also that the integral of our regulator is 1 because in the limit $\epsilon \rightarrow 0$, it tends to a Dirac delta. The prefactor $\varrho$ is here to compensate the square root of the determinant of the 2D metric (which is the Fubini–Study metric). Specifically, $\varrho(y,z) = \frac{1}{\sqrt{\det(g_{\mu \nu}(z))}}$. By expanding $g^{-1}(z)$ in power series about $z=x$, we obtain:
\begin{align}\label{eqa48}
    D^{-1}_\text{reg.}(x,x) =& \int_{\partial \Sigma} d^2z \frac{1}{\pi^2 \epsilon}\frac{e^{-\frac{|x-z|^2}{\epsilon}}}{\pi(\overline{x}-\overline{z})}\left( \text{Id}-\partial_zg(x)g^{-1}(x)(x-z)-\partial_{\overline{z}}g(x)g^{-1}(x)(\overline{x}-\overline{z})+\cdots \right)
\nonumber \\
    =& \int_{\partial \Sigma} d^2Z \frac{1}{\pi^2 \epsilon}\frac{e^{-\frac{|Z|^2}{\epsilon}}}{\overline{Z}}\left( \text{Id}-\partial_zg(x)g^{-1}(x)Z-\partial_{\overline{z}}g(x)g^{-1}(x)\overline{Z}+\cdots \right)
\end{align}
The terms in the ellipsis will tend to zero as the regulator tends to a Dirac delta. By an antisymmetry argument, the first two terms in parentheses vanish after integration. Therefore, only the third term remains. Evaluating the integral in the limit $\epsilon \rightarrow 0$ yields $D^{-1}_\text{reg.}(x,x)=-\frac{1}{\pi}\partial_{\overline{z}}g(x)g^{-1}(x)$. Thus, (\ref{eqa45}) yields
\begin{align}\label{eqa49}
    \frac{\delta S_\text{eff.}[g]}{\delta A^a_z(x)} =& \frac{\delta \ln \det(D)}{\delta A^a_z(x)} = \text{Tr}\left[ \frac{1}{\pi}\partial_{\overline{z}}g(x)g^{-1}(x)(i\tau^a) \right]
\nonumber \\
    \Rightarrow& \delta S_\text{eff.}[g] = \frac{1}{\pi}\int_{\partial \Sigma} d^2x \text{Tr}[\partial_{\overline{z}}g(x)g^{-1}(x) \delta A_z(x)]
\nonumber \\
    \Rightarrow& \delta S_\text{eff.}[g] = -\frac{1}{\pi}\int_{\partial \Sigma} d^2x \text{Tr}[\overline{C}(x) \delta A_z(x)]
\end{align}
However, this is exactly the variation of the WZW action (\ref{eqa24.2}). We can thus write:
\begin{equation}\label{eqa50}
    S_\text{eff.}[g]\approx \frac{c_A}{2}S_\text{WZW}[g] \Rightarrow \det(D) = \det(\partial_z)e^{\frac{c_A}{2}S_\text{WZW}[g]}
\end{equation}
The symbol $\approx$ indicates that this is an equality up to an additive ``constant,'' which is traduced by the most natural one, namely $\ln\det(\partial_z)$. The factor $c_A/2$ comes from the trace in the adjoint representation and our convention for the normalization of the Killing form on $\mathfrak{su}(2) \otimes_\mathbb{R} \mathbb{C} \simeq \mathfrak{sl}(2,\mathbb{C})$. Thus, the integration measure becomes:
\begin{equation}\label{eqa51}
    \mathcal{D}A_z\mathcal{D}A_{\overline{z}} = e^{\frac{c_A}{2}S_\text{WZW}[h]}\det(\partial^2)\mathcal{D}U\mathcal{D}h,\,\,\,h=g^\dagger g = \rho^2
\end{equation}
Where we have used the Polyakov-Wiegmann equation to express the sum of the two WZW actions (coming from $\ln\det(D^2)=\ln\det(D)+\ln\det(\overline{D})$), and the local counter-term $\int A_z\wedge A_{\overline{z}}$ (whose introduction is equivalent to a choice of renormalization scale) as $S_\text{WZW}[g^\dagger g]$. We can safely redefine the integration measure $\mathcal{D}h \leadsto \mathcal{D}h\times \left[\det(\partial^2)\mathcal{D}U\right]^{-1}$. Therefore, we can write the following.
\begin{align}\label{eqa52}
    Z^\text{CS}[A_z|_\partial] =& \int \mathcal{D}h\,e^{(k+\frac{c_A}{2})S_\text{WZW}[h]+\frac{ik}{2\pi}\int_{\partial\Sigma}d^2z \text{Tr}[A_z|_\partial\,(h^{-1}\partial_{\overline{z}} h)]}
\nonumber \\
    \equiv& \left\langle e^{\frac{i}{2\pi}\int_{\partial\Sigma}d^2z \sqrt{\sigma}\text{Tr}[A_z|_\partial\,\overline{\mathcal{J}}]} \right \rangle_{\text{WZW}_{k+\frac{c_A}{2}}}
\end{align}
Where $\sqrt{\sigma}$ is the determinant of the 2D Fubini-Study metric. This completes the proof that $Z^\text{3D}$ can be written using two $H_3^+$-WZW generating functionals (with $H_3^+\equiv \text{SL}(2,\mathbb{C})/\text{SU}(2)$). Alternatively, we could have used the equation $F_{z,\overline{z}}Z[A_z|_{\partial}]$ and also concluded that the partition function is a $H_3^+$-WZW generating functional \cite{MacCormack}, but this method hides under the carpet many details.\\
\indent Now, we can complete the calculation performed in the previous section, namely the semi-classical contribution to the functional integral. We can derive the equations of motion of the WZW$^+$ theory with source for the variable $\overline{J}\equiv k^{-1} \overline{\mathcal{J}}$, which are:
\begin{equation}\label{eqa53}
    i(2k+c_A)\partial_z \overline{J} = ik\partial_{\overline{z}} (\sigma^{\overline{z}z}A_z|_{\partial}) \Longrightarrow \left( \overline{J} = \frac{1}{2} \left(1+\frac{c_A}{2k}\right)^{-1} \sigma^{\overline{z}z}A_z|_{\partial}\,\,\,\&\,\,\,h^{-1}\partial_z h = -\frac{1}{2} \left(1+\frac{c_A}{2k}\right)^{-1} A_z|_\partial \right)
\end{equation}
To obtain the semi-classical approximation, we simply inject these solutions into (\ref{eqa40.2}). So we obtain:
\begin{align}\label{eqa54}
    \left\langle e^{\frac{i}{2\pi}\int_{\partial\Sigma}d^2z \text{Tr}[A_z|_\partial\,\overline{\mathcal{J}}]} \right \rangle_\text{WZW} \simeq& \left.e^{kS_\text{WZW}[h]+\frac{ik}{2\pi}\int_{\partial\Sigma}d^2z \text{Tr}[A_z|_\partial\,(h^{-1}\partial_{\overline{z}} h)]}\right|_{\overline{\mathcal{J}} =  \frac{1}{2}\left(1+\frac{c_A}{2k}\right)^{-1}\sigma^{\overline{z}z} A_z|_{\partial}}
\nonumber \\
    =& e^{\frac{i}{8\pi}\frac{k}{2}\left(1+\frac{c_A}{2 k}\right)^{-1}\int_{\partial \Sigma} d^2z \sqrt{\sigma}A^a_z|_\partial \sigma^{\overline{z}z} A^a_z|_\partial}
\end{align}
The $1/2$ factor originates from the trace in the fundamental representation of the $\text{SU}(2)$ basis. We recall (\ref{eqa40}) that we have two WZW models with opposite signs, one for $A^+$ and the other for $A^-$. Moreover, we need to choose a boundary condition for the gauge field $A_z$ that recreates the right entropy (up to small corrections). Doing the calculation, one can see that the condition $A^{a}_z|_\partial \sigma^{\overline{z} z} A^{a}_z|_\partial = \pi$ gives the same result as the usual one given in \cite{cassaniBHlectures}. This condition does not allow us to directly find the form of $A_z$, but we can nonetheless give an expression to it. For example, $A^a_z|_\partial = \alpha^a_z \sqrt{\pi}\sqrt{\sigma}^{-1/2}$ with $\alpha^{a}_z \alpha^{a}_z = 1$ gives back our condition (see the next section for a derivation of the entropy involving another $A_z|_\partial$). We introduce the change $k\leadsto \frac{1}{2} \beta k$, with 1/2 to counter the Keldysh contour:
\begin{align}\label{eqa55}
    -\ln(Z_\text{K.}[A^\pm|_\partial]) \simeq&\, \frac{\beta}{32\pi}\frac{k}{1+\frac{c_A}{\beta k}}\int d^2z\,\sqrt{\sigma}A^{+,a}_z|_\partial \sigma^{\overline{z} z} A^{+,a}_z|_\partial - \frac{\beta}{32\pi}\frac{k}{1-\frac{c_A}{\beta k}}\int d^2z\,\sqrt{\sigma}A^{-,a}_z|_\partial \sigma^{\overline{z} z} A^{-,a}_z|_\partial
\nonumber \\
    =& -\frac{\beta^2}{16\pi} - \mathcal{O}(\Lambda)
\end{align}
This is because, due to the spherical symmetry of our 3+1D metric $g_\text{Sch.}$, the 2D flat angular coordinates coincide with the 2D angular coordinates of $g_\text{Sch.}$. Note that our method is completely different from the usual one because we never used the limit $\partial \Sigma \rightarrow \infty$ in the evaluation of the boundary terms in the case of the Schwarzschild metric.\\
\indent Finally, assuming that the level $k$ of the WZW models is an integer (this is not obvious as $\frac{1}{\sqrt{\Lambda}}$ has a large value and there is no \textit{a priori} reason for it to be an integer), we find by an argument of large gauge transformation invariance that $2\beta_m\sqrt{\Lambda}^{-1}$ ought to be an integer too. As a half-unit, we could choose $\beta_m = t_\text{Pl}$, the Planck time. This is because $2\beta_m$ would then be homogeneous to the space unit of $2\ell_\text{Pl}$ hidden in the Bekenstein-Hawking entropy. However, $\beta_m=\sqrt{\Lambda}\ell_\text{Pl}t_\text{Pl}$ may be a better choice, and we choose this latter one.

\subsection*{4) Liouville gravity and holographic graviton exchange}

Knowing the form of the partition function is the first step in holographic calculations. Here, we first reduce the theory to a Liouville one upon choosing a suitable source $A_z|_\partial$, and show that the temperature and the entropy of the boundary are unchanged. Next, we calculate the effect of a massive particle on a lighter one from the cosmological boundary. Usually, the bulk is reconstructed via the dictionaries of the AdS/CFT correspondence linking the partition function in the bulk and the generating functional on the boundary \cite{Kajuri_2021}. We will choose another method, namely the worldline formalism \cite{maxfield2017view}. First, we use a Gauss decomposition of the $H_3^+$-valued field $h$ we integrate on in the correlation (\ref{eqa52}):
\begin{equation}\label{eqa8.1}
    h=\left( \begin{matrix}
        1 & \gamma \\
        0 & 1
    \end{matrix} \right)
    \left( \begin{matrix}
        e^{\varphi} & 0 \\
        0 & e^{-\varphi}
    \end{matrix} \right)
    \left( \begin{matrix}
        1 & 0 \\
        \overline{\gamma} & 1
    \end{matrix} \right)
\end{equation}
Then, the WZW action becomes the Liouville-like action:
\begin{equation}\label{eqa-Liouville-like}
    S^\pm_\text{WZW}[h] = \frac{i}{2\pi}\left(\frac{c_A}{2}\pm k\right)\int_\mathbb{C} d^2z \left(2\sqrt{\sigma} \sigma^{\mu \nu} \partial_\mu \varphi \partial_\nu \varphi + \sqrt{\sigma} \sigma^{\mu \nu}\partial_\mu \overline{\gamma} \partial_\nu \gamma e^{-2\varphi}\right)
\end{equation}
The integration measure $\mathcal{D}h$ on $h$ changes to an invariant integration measure $\mathcal{D}\varphi \mathcal{D}^2(\gamma e^{-\varphi})$. Integrating on $\gamma e^{-\varphi}$ gives a determinant $\det^{-1}(e^{-\varphi}\nabla_z\{e^{2\varphi} \nabla_{\overline{z}}[e^{-\varphi}\,\bullet]\}) \equiv \det^{-1}(D^2)$. We can rewrite it as:
\begin{equation}
    \det(D^2) = \det(D)\det(\overline{D})
\end{equation}
With $D=D_z=\nabla_z+a_z$ and $\overline{D}=D_{\overline{z}}=\nabla_{\overline{z}}-a_{\overline{z}}$, for $a_z=f^{-1}\partial_z f$, and $f=e^{\sigma^{(3)}\varphi}$. We use the result (\ref{eqa50}), namely that:
\begin{equation}
    \det(D)\det(\overline{D})=\det(\partial^2)e^{\frac{c_A}{2}S_\text{WZW}[f]}
\end{equation}
It remains to evaluate $\det(\partial^2)$. We could redefine the measure $\mathcal{D}h$ so that it absorbs it, as before. But notice that in the previous section, this was possible because $\det(D^2)$ was the determinant of the Jacobian of a change of measure. Here, this is no longer true as now $\det^{-1}(D^2)$ is not a Jacobian, its interpretation is different, and so is its treatment. To evaluate $\det(\partial^2)$, we use the fact that $e^{-2\varphi}$ expresses the distance scale on the sphere, $\varphi$ is a dilaton. Usually, this would not be the case but we recall that this is a theory of the geometry on the sphere, and so $\sigma$ is the background metric while $e^{2\varphi}\sigma$ is the true metric. Now, we can calculate $\det(\partial^2)$. We use the zeta regularization and introduce $\mu^2$, an energetic scale coming from this regularization method. Specifically, its purpose is to ensure that the argument of logarithmic quantities is dimensionless. Thus we define:
\begin{equation}
    \ln \det_\zeta(\partial^2)|_{\mu^2} \equiv \int_\mathbb{C} d^2z\left( \frac{1}{4\pi \epsilon^2}\sqrt{\sigma} +\frac{1}{24\pi}\left[-\gamma_\text{E} +\ln\left( \frac{\epsilon^2 \mu^2}{2} \right)\right]\sqrt{\sigma}R\right)+\mathcal{O}(\epsilon^2)
\end{equation}
We introduce the scales $\mu^2 = \frac{1}{l^2}e^{2\alpha\varphi+\gamma_\text{E}}$ with $l^2\rightarrow 0$ an area, and $\epsilon^2 = l^2 e^{-2b\varphi}$. With $b=\sqrt{\mp i k}$ but we will change the integration measure as $\mathcal{D}\varphi \leadsto \mathcal{D}(\sqrt{\mp i k}\varphi)$. Indeed, the kinetic term in (\ref{eqa-Liouville-like}) suggests the use of this measure for the path integral to be well-defined if we want a space-like Liouville theory. So we have:
\begin{equation}
    \ln \det_\zeta(\partial^2)|_{\mu^2}= \frac{1}{4\pi l^2}\int_\mathbb{C} d^2z\sqrt{\sigma}e^{2b\varphi} -\frac{b-\alpha}{12\pi}\int_\mathbb{C} d^2z\sqrt{\sigma}R\varphi
\end{equation}
Where $\alpha$ is a parameter we are free to choose, depending on the theory (if $\text{sgn}(\pm k)=\pm 1$). Shifting $\varphi \leadsto \varphi + \frac{1}{2b}\ln(l^2)$, the determinant $\det^{-1}_\zeta(D^2)$ is thus of the form:
\begin{equation}
    \det_\zeta{}^{-1}(D^2)|_{\mu^2} \propto e^{-\frac{1}{4\pi}\int_\mathbb{C} d^2z\sqrt{\sigma}e^{2b\varphi} +\frac{1}{12\pi}\int_\mathbb{C} d^2z\sqrt{\sigma}\left((b-\alpha) R\varphi -6c_A\partial_\mu \varphi \partial^\mu \varphi \right)}
\end{equation}
So the partition function becomes:
\begin{align}\label{eqa10.1}
    Z =& \int \mathcal{D}\varphi\mathcal{D}^2(e^{-\varphi}\gamma)\, e^{\frac{i}{2\pi}\int_\mathbb{C} d^2z \left(2\left(\frac{c_A}{2}\pm k\right)\sqrt{\sigma} \sigma^{\mu \nu} \partial_\mu \varphi \partial_\nu \varphi + k \sqrt{\sigma} \sigma^{\mu \nu}\partial_\mu \overline{\gamma} \partial_\nu \gamma e^{-2\varphi}\right)}
\nonumber \\
    \stackrel{\zeta\,\text{reg.}}{\propto}& \int \mathcal{D}(\sqrt{\mp i k}\varphi) \, e^{\frac{i}{2\pi}\int_\mathbb{C} d^2z \left(2\left(\frac{c_A}{2}\pm k-\frac{c_A}{2}\right)\sqrt{\sigma} \sigma^{\mu \nu} \partial_\mu \varphi \partial_\nu \varphi + i\frac{b-\alpha}{6} \sqrt{\sigma} R \varphi+ \frac{i}{2}\sqrt{\sigma} e^{2\sqrt{\mp i k}\varphi}\right)},\,\,\,\alpha \equiv 12\sqrt{\mp i k}+b
\nonumber \\
    \stackrel{!}{\propto}& \int \mathcal{D}\varphi\, e^{-\frac{1}{\pi}\int_\mathbb{C} d^2z \left( \sqrt{\sigma} \sigma^{\mu \nu} \partial_\mu \varphi \partial_\nu \varphi + Q \sqrt{\sigma} R \varphi+ \frac{1}{4}\sqrt{\sigma} e^{2|b|\varphi}\right)},\,\,\, |b|=1,\,\, Q\equiv |b|+\frac{1}{|b|}=2
\end{align}
The result does not depend on $\pm k$ anymore, which is what we sought with this change of variable and this definition of $b$. The background charge $Q$ has been chosen so that it is precisely $|b|+\frac{1}{|b|}=2$. Moreover, the central charge of the theory $c=1+6Q^2$ has the smallest possible value ($c=25$). At least, all this is true if we forget the source term of the WZW model. To take it into account, notice that due to the Gauss decomposition we have $A^{(z)}_z|_\partial \overline{\mathcal{J}}{}^{(z)} = 0$ and $A^{(\overline{z})}_z|_\partial \overline{\mathcal{J}}{}^{(\overline{z})} \neq 0$. Upon choosing $A^{(\overline{z})}_z|_\partial=0$, which seems logical, we obtain that only the $A^{(3)}_z$ contribution will remain. We choose:
\begin{equation}\label{eqa11.1}
    A^{(3)}_z|_\partial(z) = \frac{2}{\rho}a^{(3)}_z\sum_{i=1}^N \alpha_i\left[\frac{\rho}{z-z_i}+\frac{1}{2i\pi} \oint_{\partial D(z)} dw \frac{1}{w-z_i}\right]
\end{equation}
With $a^{(3)}_z=\sqrt{\sigma}^{-1}$, and $\rho = \eta \sqrt{\sigma}^{-1/2}$ is the radius of $D(z)$, a disk centered on $z$, which will have all its significance when we calculate the entropy. With (\ref{eqa11.1}), and after some calculations, we obtain:
\begin{equation}\label{eqa12.1}
    \int d^2z \,\text{tr}[A_z|_\partial \overline{\mathcal{J}}] = 4\pi \sum_{i=1}^N \alpha_i\varphi(z_i)
\end{equation}
The positions $z_i$ have to be evenly scattered on the boundary/horizon, which is a sphere in the best case. But because we are using a stereographic projection to be able to use the coordinates $z$, the radius $\rho$ has to grow with distance to the center of the projection. This is the reason behind the definition $\rho = \eta \sqrt{\sigma}^{-1/2}$ ($\sqrt{\sigma}^{-1}$ would be a growing area, so we take the square root). The gauge field component $A^3_z|_\partial(z)$ has uncommon properties because it has poles, and a discretization of the boundary too due to the second term in brackets in (\ref{eqa11.1}). The correlation functions we find with $Z_\text{K.}$ are the square of correlations of a Liouville theory (we choose $A^{(3),+}_z|_\partial=A^{(3),-}_z|_\partial$). But the two-point function is, for example:
\begin{equation}\label{eqa13.1}
    \left[\langle e^{\alpha_i \varphi(z_i)}e^{\alpha_j \varphi(z_j)} \rangle_\text{L.}\right]^2 \sim \frac{1}{|z_i-z_j|^{4\Delta_{ij}}}
\end{equation}
Where $\Delta_{ij}=\Delta_i+\Delta_j$, the sum of the conformal dimensions of $\alpha_i$ and $\alpha_j$. To have a graviton propagator, we need to have $\Delta_i+\Delta_j=2$. Doing the same reasoning for all correlation functions, we deduce that the only solution is to have $\Delta_i=1$ for all $i$. This is possible if $\alpha_i=1$ because $\Delta_i = \alpha_i(Q-\alpha_i)$, and $Q=2$. Note that $\alpha_i = \frac{Q}{2}$, so it is possible to add an imaginary part, a ``momentum'', and as we will see after, it corresponds to particles. \\
\indent We focus now on the entropy for the Schwarzschild black hole, hoping that it will not change under our definition of $A^{(3)}$. Upon decomposing $h = h_\text{cl.}\mathfrak{h}$ in the $+$ side of the WZW model (\ref{eqa52}), we obtain:
\begin{align}\label{eqa14.1}
    Z[A_z|_\partial] =& \int \mathcal{D}h\,e^{\left( k+\frac{c_A}{2} \right)S_\text{WZW}[h]+\frac{ik}{2\pi}\int_\partial d^2z\,\text{tr}[A_z|_\partial (h^{-1}\partial_{\overline{z}}h)]}
\nonumber \\
    \leadsto& \int \mathcal{D}(h_\text{cl.} \mathfrak{h})\,e^{\left( k+\frac{c_A}{2} \right)\left[S_\text{WZW}[h_\text{cl.}]+S_\text{WZW}[\mathfrak{h}]-\frac{i}{2\pi}\int d^2z \,\text{tr}[h_\text{cl.}^{-1} \partial_z h_\text{cl.}\,\mathfrak{h}^{-1}\partial_{\overline{z}} \mathfrak{h}]\right]}e^{\frac{ik}{2\pi}\int_\partial d^2z\,\text{tr}[A_z|_\partial (h_\text{cl.}^{-1}\partial_{\overline{z}}h_\text{cl.}+\mathfrak{h}^{-1} \partial_{\overline{z}} \mathfrak{h})]}
\nonumber \\
    \leadsto& \det(h_\text{cl.})e^{\left( k+\frac{c_A}{2} \right)S_\text{WZW}[h_\text{cl.}]+\frac{ik}{2\pi}\int_\partial d^2z\,\text{tr}[A_z|_\partial (h_\text{cl.}^{-1}\partial_{\overline{z}}h_\text{cl.})]} \int \mathcal{D}\mathfrak{h}\,e^{\left( k+\frac{c_A}{2} \right)S_\text{WZW}[\mathfrak{h}]+\frac{ik}{4\pi}\int_\partial d^2z\,\text{tr}[A_z|_\partial (\mathfrak{h}^{-1} \partial_{\overline{z}} \mathfrak{h})]}
\end{align}
Where we used the Polyakov-Wiegmann identity in the second line. Because $A_z|_\partial$ is traceless, we have $\det(h_\text{cl.})=1$. Now, we recall that the solution to the classical equations of motion for $J^{\overline{z}}_\text{cl.} = \overline{J}_\text{cl.} \equiv k^{-1} \overline{\mathcal{J}}_\text{cl.}$ are:
\begin{equation}\label{eqa15.1}
    i(2k+c_A)\partial_z \overline{J}_\text{cl.} = ik\partial_{\overline{z}} (\sigma^{\overline{z}z}A_z|_{\partial}) \Longrightarrow \overline{J}_\text{cl.} = \frac{1}{2} \left(1+\frac{c_A}{2k}\right)^{-1} \sigma^{\overline{z}z}A_z|_{\partial}
\end{equation}
So that the classical action is $S_\text{WZW}[h_\text{cl.}] = \frac{i}{8\pi}\frac{k}{2}\left(1+\frac{c_A}{2 k}\right)^{-1}\int_{\mathbb{C}} d^2z \sqrt{\sigma}A^a_z|_\partial \sigma^{\overline{z}z} A^a_z|_\partial$. As we choose $A^{(\overline{z})}_z|_\partial=0$, it remains to specify $A^{(z)}_z|_\partial$. We find that an appropriate choice is:
\begin{equation}\label{eqa16.1}
    A^{(z)}_z|_\partial (z)= \frac{2}{\rho} a^{(z)}_z\sum_{i=1}^N \frac{\sqrt{2-\alpha_i}}{2i\pi} \oint_{\partial D(z)} dw \frac{1}{w-z_i}
\end{equation}
With $a^{(z)}_z=\sqrt{\sigma}^{-1}$. Of course, all the $\alpha_i$ are equal to $1$. This form (\ref{eqa16.1}) is interesting because it contains a discretization of the boundary too. So if $D(z)$ does not contain any $z_i$, we obtain $0$, which is interpreted as a coordinate singularity in $A_{(z)}^z$ (the inverse of (\ref{eqa16.1})). Thus, we have a singularity in the $z$ component of the radial gauge field at some points and a cancellation of the $z$ component of the angular gauge field in between. Because the poles vanish when integrated over the whole stereographic projection, the $+$ side of the classical WZW action is then, in Euclidean signature:
\begin{align}\label{eqa17.1}
    &\int d^2z \sqrt{\sigma}\left(A^{(z)}_z|_\partial (z)  \sigma^{\overline{z} z} A^{(z)}_z|_\partial (z)+A^{(3)}_z|_\partial (z) \sigma^{\overline{z} z} A^{(3)}_z|_\partial (z) \right) \stackrel{!}{=} \varrho\frac{2\beta^2_s}{\eta^2 \pi}
\end{align}
Where $\varrho$ is the ratio between the area $\pi \eta^2$ covered by $D(z)$ on the spherical boundary, and the elementary area $4\ell_\text{Pl}^2$. $\beta_s$ is the inverse of the Hawking temperature. We have to choose $\eta$ so that it matches $2\beta_m$. As we can see, posing $\eta=2\beta_m$ makes $A^{(z)}_z|_\partial=0$ most of the time. But these non-zero patches on the boundary carry a weight of $\frac{1}{\eta}$ so each of them effectively represents an area of $A = 4\ell_\text{Pl}^2$. With this, the total Euclidean classical action is found to be:
\begin{equation}\label{eqa18.1}
    -S^\pm[h_\text{cl.}] = \frac{1}{32\pi}\frac{\beta_s^2}{2} \left(\frac{\beta k}{1+\frac{c_A}{\beta k}}-\frac{\beta k}{1-\frac{c_A}{\beta k}} \right) = - \frac{\beta_s^2}{16\pi}-\mathcal{O}(\Lambda)
\end{equation}
Note that we have changed $2\times (2\beta_m)$ into $\beta$, the Euclidean temperature of the system (which almost coincides with $\beta_s$), because $2$ comes from the Keldysh contour that we suppress to use a thermal one, and $2\beta_m \leadsto \beta$ because $2\beta_m$ was our unit of time. The result (\ref{eqa18.1}) is the same as \cite{cassaniBHlectures} up to small corrections, and gives the Bekenstein-Hawking entropy, and the Hawking temperature. Thus, we conclude this derivation by the fact that our choice of geometry (\ref{eqa11.1}) and (\ref{eqa16.1}) does not change the entropy and the temperature of the system because the total action is the same, up to quantum corrections coming from the path integral of (\ref{eqa14.1}). \\
\indent Now, we introduce point particles in the theory (\ref{eqa10.1}), one of mass $M>0$ and one of mass $0<m\ll M$. The mass $M$ is at the center of the coordinate system and describes a Schwarzschild metric in which $m$ moves. The idea is to use the WKB approximation of the following partition function in the variable $\Gamma$, the trajectory of the particle of mass $m$:
\begin{equation}\label{eqa18.2}
    Z=\int \mathcal{D}g\mathcal{D}\Gamma\,e^{iS_\text{EH}[g]+iS_\text{W.L.}[g,\Gamma]}
\end{equation}
Notice that, if we had added a field action with source from the beginning of part II.1), we would have obtained a 3D propagator due to the Keldysh contour. Because we can write a 3D propagator on space-like intervals $\ll 1$ as a function of $\sqrt{\gamma_{ij}\dot{x}{}^i \dot{x}{}^j}$ (the signature of $\gamma_{ij}$ is $(+,+,+)$), we obtain that to introduce matter, only the projection onto the hypersurface $t=0$ of this quantity is useful. For point matter, we need to use the exponentiated action $e^{im\int_\Gamma ds}$, which is nothing but the exponential of a relativistic point particle action projected onto the hypersurface $t=0$. Thus, it is sufficient to focus on the following:
\begin{align}\label{eqa19.1}
    e^{im\int_0^{2\beta_m}dt \sqrt{\gamma_{ij} \dot{x}^i \dot{x}^j}} \leadsto e^{i\frac{m}{2\sqrt{\Lambda}}\int_0^{2\beta_m}dt \sqrt{\frac{1}{2}\text{tr}[(\sigma^a A^a_i \dot{x}^i)^2]}}
\end{align}
This roughly looks like a Wilson line, so we need to follow the standard procedure for finding the wave-functional of a WZW model with insertions of Wilson lines \cite{CSKnotshell,gawedzki1999conformal,LABASTIDA1989214}. Specifically, we need to take the derivative on $A_r$ to find the wave-functional, which is our partition function. But our derivative is:
\begin{equation}\label{eqa20.1}
    \frac{\delta}{\delta A_r}\left(i\frac{m}{2\sqrt{\Lambda}}\int_0^{2\beta_m}dt \sqrt{\frac{1}{2}\text{tr}[(A_i \dot{x}^i)^2]}\right)=\frac{im}{2\sqrt{\Lambda}}\int_0^{2\beta_m}dt \frac{\dot{x}^r \dot{x}^iA_i}{\sqrt{\frac{1}{2}\text{tr}[(A_i \dot{x}^i)^2]}}\delta^{(2)}(z-w)\delta(r-r')
\end{equation}
We impose that $A^{(\overline{z})}_i x^i=0$, implying $x^{\overline{z}}=0$, so that the trace is $\text{tr}[(A_i \dot{x}^i)^2] = 2 (A^{(3)}_i \dot{x}{}^i)^2$. This is true for the $H_3^+$ basis that may be built from the Pauli matrices. We can see that taking this assumption into account, we obtain the expectation value of a Wilson line. Taking the normalized trace of the Wilson line we obtain (the results for $A^+_r$ and $A^-_r$ are the same, hence $A_r$ without $\pm$):
\begin{align}\label{eqa21.1}
    \frac{1}{2}\text{Tr}\left[e^{i\frac{m}{2\sqrt{\Lambda}}\int_0^{2\beta_m}dt \dot{x}^r \sigma^{(3)} \int_w^zdy\,A_z}\right] =& \frac{1}{2}\text{Tr}\left[e^{im \dot{x}^r \sigma^{(3)} \int_w^zdy\,A_z}\right]
\nonumber \\
    =& e^{im \dot{x}^r (\varphi(z)-\varphi(w))}
\end{align}
Where we used the expression of $A_z$, which is $A_z = h^{-1}\partial_z h$, and injected the Gauss decomposition (\ref{eqa8.1}) in it. We could use the first line as in \cite{Be_ken_2019} and insert it in the WZW model, but it would lead to a perturbative result in powers of the source of the model. Introducing in the same manner $\mathcal{N}$ point particles we obtain the sourced partition function:
\begin{align}\label{eqa22.1}
    Z_\text{K.}[\alpha_i | z_i | \mathcal{N}] =&  \left[ \left\langle \prod_{n=1}^{\mathcal{N}}\left(e^{2P_n\varphi(z_{i_n})}e^{2\overline{P}_n\varphi(z_{j_n})}\right)\prod_{k\notin \{i\}\cup\{j\}} e^{2\alpha_k \varphi(z_k)} \right\rangle_\text{L.} \right]^2
\end{align}
With $P_n = \alpha_{i_n} + \frac{i}{2}m_n\dot{x}_n^r = \frac{Q}{2} + \frac{i}{2}m_n\dot{x}_n^r$. As we can see, the ``momenta'' of the vertex operators are truly momenta. We can interpret this as particles leaving the points $z_{j_n}$, and going towards $z_{i_n}$. We can go even further by completely ignoring (\ref{eqa19.1}), and directly modify (\ref{eqa11.1}) to include the momenta in the poles. The two methods are completely equivalent and we can see that the presence of point particles is encoded into the geometry of the boundary. Evaluating the correlator (\ref{eqa22.1}) is a daunting task but fortunately, when $\mathcal{N}=1$, there exists an estimate upper bound found in \cite{Kupiainen_Rhodes_Vargas}. The setup is as follows. Let $\{z_i\}_{i\neq 1,2}$ be $N-2$ points on the Riemann sphere such that their direct neighboring points are separated from them by at least a radius $\delta$. Then define $z_1,z_2 \in \overline{\mathbb{C}}  \setminus \bigcup_{i\neq 1,2} D(z_i,\delta)$, where $D(z_i,\delta)$ is a disk centered at $z_i$ and of radius $\delta$. Let the weights $\beta_1,\beta_2$ and $\alpha_k$ be such that:
\begin{align}\label{eqa23.1}
    \beta_1+\beta_2 > \frac{Q}{2} \,\,\,\& \,\,\, \sum_{i\neq 1,2} \alpha_k > \frac{Q}{2}
\end{align}
Moreover, we ought to have $\Re[\alpha_k] \leq  \frac{Q}{2}$, and the same for $\beta_{1,2}$. Upon summing up these two conditions, we arrive at the well-known Seiberg bound $\beta_1 + \beta_2 + \sum_k \alpha_k > Q$, expressing the condition for the correlation function to exist. Notice that because $\alpha_k=\Re[\beta_{1,2}]=1$, and because our horizon has an area greater than $\ell_\text{Pl}^2=1$, we are in these bounds. Then, we have the upper bound:
\begin{equation}\label{eqa24.1.0}
    \left\langle e^{2\beta_1\varphi(z_{1})}e^{2\beta_2\varphi(z_{2})}\prod_{k\neq 1,2} e^{2\alpha_k \varphi(z_k)} \right\rangle_{\epsilon,\text{L.}} \leq C_\delta \,\sigma(z_3)^{\Delta_{\alpha_4}} |z_1-z_2|_\epsilon^{2\frac{Q^2}{4}-2\Delta_{\beta_1}-2\Delta_{\beta_2}}|\ln(|z_1-z_2|_\epsilon)|^{-\frac{3}{2}}
\end{equation}
Where $|z_1-z_2|_\epsilon = \max(\epsilon,|z_1-z_2|)$, for a parameter $\epsilon \rightarrow 0$. The coefficient $C_\delta$ contains all other contributions coming from the correlator. This estimation comes from the fusion rule of the vertex operators, which is believed to be trusted in the case where $z_1 \rightarrow z_2$. But, recall that the position $z_1$ is nothing but $z_2+\eta$, where $\eta$ is a parameter $\ll1$. This is because $z_1$ and $z_2$ are the projections onto the Riemann sphere of the positions at $t=0$ and $t=2\beta_m\ll 1$. We recall that the left-hand side of (\ref{eqa24.1.0}) is the WZW expectation value of a Wilson line. To extract the information of the gravitational imprint of the mass $m$ on the boundary, we are interested in the ratio between this Wilson line, and the case where there is no mass $m$. Assuming $\epsilon < |z_1-z_2|\ll 1$, expression (\ref{eqa22.1}) reduces to, when $\mathcal{N}=1$:
\begin{align}\label{eqa25.1.0}
    \frac{Z_\text{K.}[\alpha_i | z_i | 1]}{Z_\text{K.}[\alpha_i | z_i | 0]} \sim |z_1-z_2|^{-4(\Delta_{\beta_1}+\Delta_{\beta_2}-\Delta_{\alpha_1}-\Delta_{\alpha_2})} = |z_1-z_2|^{-2 m^2 (\dot{x}^r)^2}
\end{align}
This expression is not very enlightening because $z_1-z_2$ is a position on the complex plane rather than on the sphere. Because our boundary has, at best, a spherical shape, we transform $z_1-z_2 = r_\Lambda e^{i\phi}\tan\left( \frac{\theta}{2} \right)$ to have a coordinate on the boundary without stereographic projection (with $r_\Lambda = \sqrt{\frac{3}{\Lambda}}$). Taking the limit of coincident points, we obtain:
\begin{align}\label{eqa26.1.0}
    Z_\text{K.}[\alpha_i | z_i | 1]\sim \left( \frac{4}{r^2_\Lambda \theta^2} \right)^{m^2 (\dot{x}^r)^2}\left( 1-\frac{1}{6}m^2 (\dot{x}^r)^2 \theta^2 + \mathcal{O}(\theta^4) \right)Z_\text{K.}[\alpha_i | z_i | 0]
\end{align}
This expansion resembles the typical power series in the variable $z$ on the cylinder we can find in the treatment of the four-point correlator in the context of the AdS/CFT \cite{Kaplan}. Upon maybe forcing the analogy because $\theta$ is our variable and $z$ is the variable in AdS/CFT, we are tempted to interpret the fact that $\theta$ appears in even powers as a manifestation of an exchange of virtual gravitons. But then one could ask: ``What is the significance of each term?'' The fact that at the next-to-lowest order, we have $m^2 \dot{x}^2 \theta^2 \equiv p^2 \theta^2$ and not $p^2 \theta^2 P^2$ for $P$ the static limit of the momentum of the heavy particle, can be interpreted as follows. We have taken the lighter point particle as being in the Schwarzschild metric induced by the heavier one. But then, we do not expect a term of the form $p^2 \theta^2 P^2$, the counterpart of $V_{\mu \nu}(p) G^{\mu \nu \alpha \beta}(q) V_{\alpha \beta}(P)$ ($V$ the vertex rule and $G$ the graviton propagator) because it would mean two free particles in \textit{flat} space exchanging a virtual graviton, and this is not the case. We are in curved space, as expressed more explicitly by the purely radial velocity $\dot{x}^r = \left( 1-\frac{r_s}{r} \right)\sqrt{\frac{r_s}{r}}$, and the presence of the heavy mass $M$ is reflected by this velocity only. So our virtual graviton exchange $\theta^2$ is indeed between the two probes but the theory is blind to the presence of the heavier. Furthermore, when calculating the term in $\theta^4$ coming from the exchange of two gravitons, we obtain something of the form $\sim p^2 \theta^4 + p^4 \theta^4$. In light of this explanation, it is easy to interpret these two terms: the first is a vertex with two scalar legs and two gravitons legs, with two gravitons propagators while the second is constituted by two vertices and two gravitons, the same as $p^2 \theta^2$ but squared. In a sense, in $p^{2k}$, $k$ counts the number of vertices puncturing the light mass' trajectory. As we can see, there should be a one-to-one correspondence between our holographic terms (\ref{eqa26.1.0}) and their perturbative quantum gravity counterparts.

\section*{Discussion and conclusion}

Although we performed approximations to find (\ref{eqa31}), it seems that the approximated transition amplitude from one 3-metric to another with fixed boundary conditions is at least close to being a heat kernel in the superspace. With the three-dimensional Einstein-Hilbert action with cosmological constant as inertia in this space, the Keldysh partition function corresponds to its diagonal and it seems that all the possible generalized Seeley-DeWitt coefficients but the zeroth one vanish on this diagonal. Thus, it appears that the full transition amplitude is indeed the diagonal of a heat kernel.\\
\indent With the time parameter defined by (\ref{eqa35}) we can try to interpret the notion of ``time'' in this context. For the flat coordinates $(r,\theta,\varphi)$ there is a horizon in the de Sitter metric. This implies that there is a region in which $\frac{1}{(f^H)^2} = |g^{00}|$ is divergent. Thus, while there is one unit of time passing in the flat coordinates, there are $\frac{1}{f^H}$ units passing at the horizon for these very coordinates (the flat coordinates can be interpreted as a Minkowskian $\mathbb{R}^{1,3}$ superimposed to our foliation $\mathcal{M}$). However, at the horizon, and for the coordinates near the horizon, only one time lapse has passed. The quantity $\xi$ in (\ref{eqa36}) is precisely the factor of this gravitational time dilation. Until now, nothing new except (\ref{eqa36}) which traduces a collection of reference frames, each of which evolves from $t=0$ to $t=\beta_m$ from their point of view. We could try to mathematize this by saying that we have a state $|\Psi\rangle =\sum_{x\in \Sigma}\Psi(t(x),x)|x\rangle$, but the notations are confusing as we have to think of $x$ as being a point on a topological space and \textit{not} as a coordinate. All this is due to the parametrization with flat coordinates. Indeed, if we had expressed the Schwarzschild metric in the Lemaitre ones, then the conclusion would have been that everything is evolving at the same ``rate''. Technically $\alpha \beta_m$ is not the time \textit{per se}, but rather a parameter chosen only to make the space-like hypersurfaces evolve in a given manner. The link between this global parameter on the worldline in the superspace, and its local effect on the worldlines in the foliation is expressed through (\ref{eqa36}). \\
\indent Our theory exhibits a dimensional reduction from 3+1D to 2D. It is interesting because it coincides with the theoretical indications that in some sense, Quantum Gravity may become two-dimensional at high energies. The interpretation of our Keldysh partition function with constant boundary conditions is that at $t=0$, a boundary is embedded in the initial space-like hypersurface. Subsequently, all the information of the theory, and most importantly, the future information, is encoded in this partition function. This means that if the boundary evaporates, then the partition function will also indicate this. And it does, because when naively Wick-rotating the partition function, we obtain the Bekenstein-Hawking entropy from the boundary temperature. Moreover, we can reconstruct the bulk gravitational interaction with the boundary and conclude that upon using a Liouville gravity theory, there is an exchange of virtual gravitons between two massive probes encoded on the boundary. The fact that there is \textit{a priori} no obstruction to construct a quantum theory of 3+1D gravity's initial conditions describing the full theory means that at first sight, the theory is unitary.\\
\indent We conclude that in light of this approach of superspace worldline formalism, Quantum Gravity may be closer to regular Quantum Mechanics than Quantum Field Theory.

\section*{Acknowledgments}


The author is grateful to the anonymous referees for their comments on the previous versions of this manuscript, which were valuable for the final writing.

\divider
\section*{References}
\nocite{*}
\bibliographystyle{iopart-num}
\bibliography{Bibliography_revision}

\appendix
\section*{Appendix A}
\setcounter{section}{1}

In our reformulation of the Keldysh partition function in terms of gauge fields, we encounter a variation of an effective action of the following form:
\begin{equation}\label{eqa24.1}
    \delta S_\text{eff.}[g] = -\frac{1}{\pi}\int_{\partial \Sigma} d^2x \text{Tr}[\overline{C}(x) \delta A_z(x)]
\end{equation}
Where $A_z$ is a two-dimensional gauge field expressed as $A_z = -\partial_z gg^{-1}$, and $\overline{C} = -\partial_{\overline{z}} gg^{-1}$, where $g$ is a $G^\mathbb{C}$-valued field ($G$ is the gauge group of the theory). To interpret this quantity, we must examine the $G$-WZW model. Indeed, it turns out that this variation is precisely the variation of the WZW action (see \cite{gawedzki1999conformal} for more details about the interpretation of such action):
\begin{align}\label{eqa24.2}
    S_\text{WZW}[g]=\frac{1}{2\pi} \int_{\partial \Sigma} \text{Tr}[\partial_z g \wedge\partial_{\overline{z}} g^{-1}]+\frac{i}{12\pi}\int_{\mathbb{B}}\text{Tr}[(g^{-1}dg)^{\wedge 3}]
\end{align}
The second term is topological and called the ``Wess-Zumino'' term. Its integration domain is any space $\mathbb{B}$ such that $\partial \mathbb{B}=\partial \Sigma$. Its value is proportional to $2\pi i$, regardless of $\mathbb{B}$, and for an integer proportionality constant. The variation of this action is \cite{EberhardtWZW}:
\begin{align}\label{eqa24.3}
    \delta S_\text{WZW}[g] =& \frac{1}{\pi}\int_{\partial \Sigma} d^2x \text{Tr}[\delta gg^{-1} \partial_{\overline{z}}(\partial_z g g^{-1})]
\nonumber \\
    =&\frac{1}{\pi}\int_{\partial \Sigma} d^2x\text{Tr}[\delta gg^{-1} \left\{\partial_z(\partial_{\overline{z}} g g^{-1})+\partial_{\overline{z}}gg^{-1}\partial_z gg^{-1}-\partial_z gg^{-1}\partial_{\overline{z}} gg^{-1} \right\}]
\nonumber \\
    =&-\frac{1}{\pi}\int_{\partial \Sigma} d^2x\text{Tr}[\delta gg^{-1} \left\{\partial_z\overline{C}+[A_z;\overline{C}] \right\}],\,\,\,\overline{C} \equiv -\partial_{\overline{z}}gg^{-1}
\nonumber \\
    =&-\frac{1}{\pi}\int_{\partial \Sigma} d^2x\text{Tr}[\delta gg^{-1} D_z\overline{C}]
\nonumber \\
    =&-\frac{1}{\pi}\int_{\partial \Sigma} d^2x\text{Tr}[\overline{C}\delta A_z]
\end{align}
For the last line, we simply used the same procedure as above, but with $\delta$ and $\partial_z$, and permuted some terms because we are dealing with a trace. Because the two variations are the same, $S_\text{eff.}$ and $S_\text{WZW}$ differ from a constant; therefore, we can write $S_\text{eff.}=S_\text{WZW}+c^\text{te}$. This is the equality we were looking for in this subsection, and it will be of the most importance in the end.

\end{document}